\documentclass[10pt,conference]{IEEEtran}

\usepackage{amsmath,amsthm,amssymb}
\usepackage[backend=biber, hyperref=false, style=ieee, sorting=none, doi=false, isbn=false, url=false]{biblatex}

\usepackage{graphicx}
\usepackage{subcaption}
\usepackage{color}

\newtheorem{definition}{Definition}
\newtheorem{claim}{Claim}

\usepackage{scalerel}
\usepackage{tikz}
\usetikzlibrary{svg.path}

\definecolor{orcidlogocol}{HTML}{A6CE39}
\tikzset{
  orcidlogo/.pic={
    \fill[orcidlogocol] svg{M256,128c0,70.7-57.3,128-128,128C57.3,256,0,198.7,0,128C0,57.3,57.3,0,128,0C198.7,0,256,57.3,256,128z};
    \fill[white] svg{M86.3,186.2H70.9V79.1h15.4v48.4V186.2z}
                 svg{M108.9,79.1h41.6c39.6,0,57,28.3,57,53.6c0,27.5-21.5,53.6-56.8,53.6h-41.8V79.1z M124.3,172.4h24.5c34.9,0,42.9-26.5,42.9-39.7c0-21.5-13.7-39.7-43.7-39.7h-23.7V172.4z}
                 svg{M88.7,56.8c0,5.5-4.5,10.1-10.1,10.1c-5.6,0-10.1-4.6-10.1-10.1c0-5.6,4.5-10.1,10.1-10.1C84.2,46.7,88.7,51.3,88.7,56.8z};
  }
}

\newcommand\orcidicon[1]{\href{https://orcid.org/#1}{\mbox{\scalerel*{
\begin{tikzpicture}[yscale=-1,transform shape]
\pic{orcidlogo};
\end{tikzpicture}
}{|}}}}

\newcommand{\papername}{Technical Report: Insider-Resistant Context-Based Pairing for\ Multimodality Sleep Apnea Test}

\usepackage{hyperref}

\begin{document}

\title{\papername}
\author{\IEEEauthorblockN{Yao Zheng\IEEEauthorrefmark{1} \orcidicon{0000-0003-2820-1034} ~~
Shekh Md Mahmudul Islam\IEEEauthorrefmark{1} ~~
Yanjun Pan\IEEEauthorrefmark{2} ~~Marionne Millan\IEEEauthorrefmark{1} ~~Samson Aggelopoulos\IEEEauthorrefmark{1}\\
Brian Lu\IEEEauthorrefmark{1} ~~Alvin Yang\IEEEauthorrefmark{1} ~~Thomas Yang\IEEEauthorrefmark{1} ~~Stephanie Aelmore\IEEEauthorrefmark{1} ~~Willy Chang\IEEEauthorrefmark{1}\\ Alana Power\IEEEauthorrefmark{1} ~~Ming Li\IEEEauthorrefmark{2} ~~Olga Bori{\'{c}}–Lubecke\IEEEauthorrefmark{1} ~~Victor Lubecke\IEEEauthorrefmark{1} ~~Wenhai Sun\IEEEauthorrefmark{3}}
\IEEEauthorblockA{\IEEEauthorrefmark{1}University of Hawai`i at M\=anoa, USA
~~\IEEEauthorrefmark{2}University of Arizona, USA
~~\IEEEauthorrefmark{3}Purdue University, USA\\
Email: \IEEEauthorrefmark{1}\{yao.zheng,shekh,marionne,saggelop,lubrian,ayang27,thomasy4,saelmore,changw,afpower,olgabl,lubecke\}@hawaii.edu \\ \IEEEauthorrefmark{2} \{yanjunpan,lim\}@email.arizona.edu \quad \IEEEauthorrefmark{3}whsun@purdue.edu}}

\maketitle 

\begin{abstract}
The increasingly sophisticated at-home screening systems for obstructive sleep apnea (OSA), integrated with both contactless and contact-based sensing modalities, bring convenience and reliability to remote chronic disease management. However, the device pairing processes between system components are vulnerable to wireless exploitation from a non-compliant user wishing to manipulate the test results. This work presents SIENNA, an insider-resistant context-based pairing protocol. SIENNA leverages JADE-ICA to uniquely identify a user's respiration pattern within a multi-person environment and fuzzy commitment for automatic device pairing, while using friendly jamming technique to prevent an insider with knowledge of respiration patterns from acquiring the pairing key. Our analysis and test results show that SIENNA can achieve reliable ($>$ 90\% success rate) device pairing under a noisy environment and is robust against the attacker with full knowledge of the context information.



\end{abstract}



\maketitle
\IEEEdisplaynontitleabstractindextext
\IEEEpeerreviewmaketitle

\ifCLASSOPTIONcompsoc

\IEEEraisesectionheading{\section{Introduction}\label{sec:introduction}}
\else
\section{Introduction}
\label{sec:introduction}
\fi
Over twenty-five million adults in the US suffer from obstructive sleep apnea (OSA), an airway muscle-related breathing condition that involuntarily causes respiratory cessations during sleep. Poor treatment can lead to excessive daytime fatigue, high blood pressure, cardio-metabolic conditions, along with a myriad of health problems \cite{baboli_good_2015}. A traditional diagnostic procedure, known as polysomnography (PSG), requires the patient to be in a laboratory overnight with instruments of multiple sensors/electrodes to track various sleep-related physiological parameters. However, PSG is highly obtrusive, expensive, and scarce.

At-home OSA monitoring systems leverage contactless and/or contact-based sensing technologies to monitor respiratory symptoms related to OSA. They allow users to conduct self-administered tests prescribed by their doctors and are considered economical alternatives for PSG. However, At-home OSA tests are subject to test fraud, as several professions within the patient population are deeply concerned that positive OSA test results will jeopardize their careers. As a result, an OSA patient may exploit the unsupervised at-home environment to manipulate the OSA test results. Specifically, the device pairing processes between the OSA screening system components are often the target of eavesdropping and spoofing from a non-compliant user.

To combat these malicious behaviors, we introduce SIENNA: in\textbf{SI}der r\textbf{E}sista\textbf{N}t co\textbf{N}text-based p\textbf{A}iring for unobtrusive at-home OSA screening. SIENNA works with a multi-modality OSA screening system consisting of one data aggregate, e.g., the user's mobile phone, and two sensing modalities, e.g.,  a respiratory belt and a physiological radar monitoring system (PRMS). It leverages the respiration patterns collected by the respiratory belt to allow automatic pairing between the PRMS and the phone. The design of SIENNA uses a novel combination of JADE-ICA \cite{rutledge_independent_2013}, fuzzy commitment, \cite{JuelsFuzzyCommitmentScheme1999}, and friendly jamming \cite{AroraDialogCodesSecure2009,GollakotaPhysicalLayerWireless2011, MelcherIJamChannelRandomization2020}. The JADE-ICA allows the PRMS to identify the unique patterns of a person's breathing from a multi-person environment. The fuzzy commitment leverages the user's breathing patterns to establish a shared secret key between the PRMS and the mobile phone. And the friendly jamming prevents insiders, e.g., a non-compliant and unsupervised user with knowledge of the breathing patterns, from learning the security key.

We formally analyzed the security of SIENNA based on the attacker's knowledge of the context information, and implemented a laboratory prototype consisting of a mmWave PRMS (implemented with SDR and mmWave radio heads), a wireless respiratory belt, and a Android-based OSA app. We conducted an evaluation consisting of 20 subjects spanning over one month. The results show that SIENNA achieves reliable device pairing within a noisy at-home environment with multiple free moving persons in the background. It also prevents unauthorized receivers from retrieving the secret key, regardless of their locations or knowledge of the user's respiration patterns.


\section{Preliminary}
\label{sec:background}
\begin{figure}[t]
\begin{subfigure}[t]{0.23\textwidth}
\includegraphics[width=\textwidth]{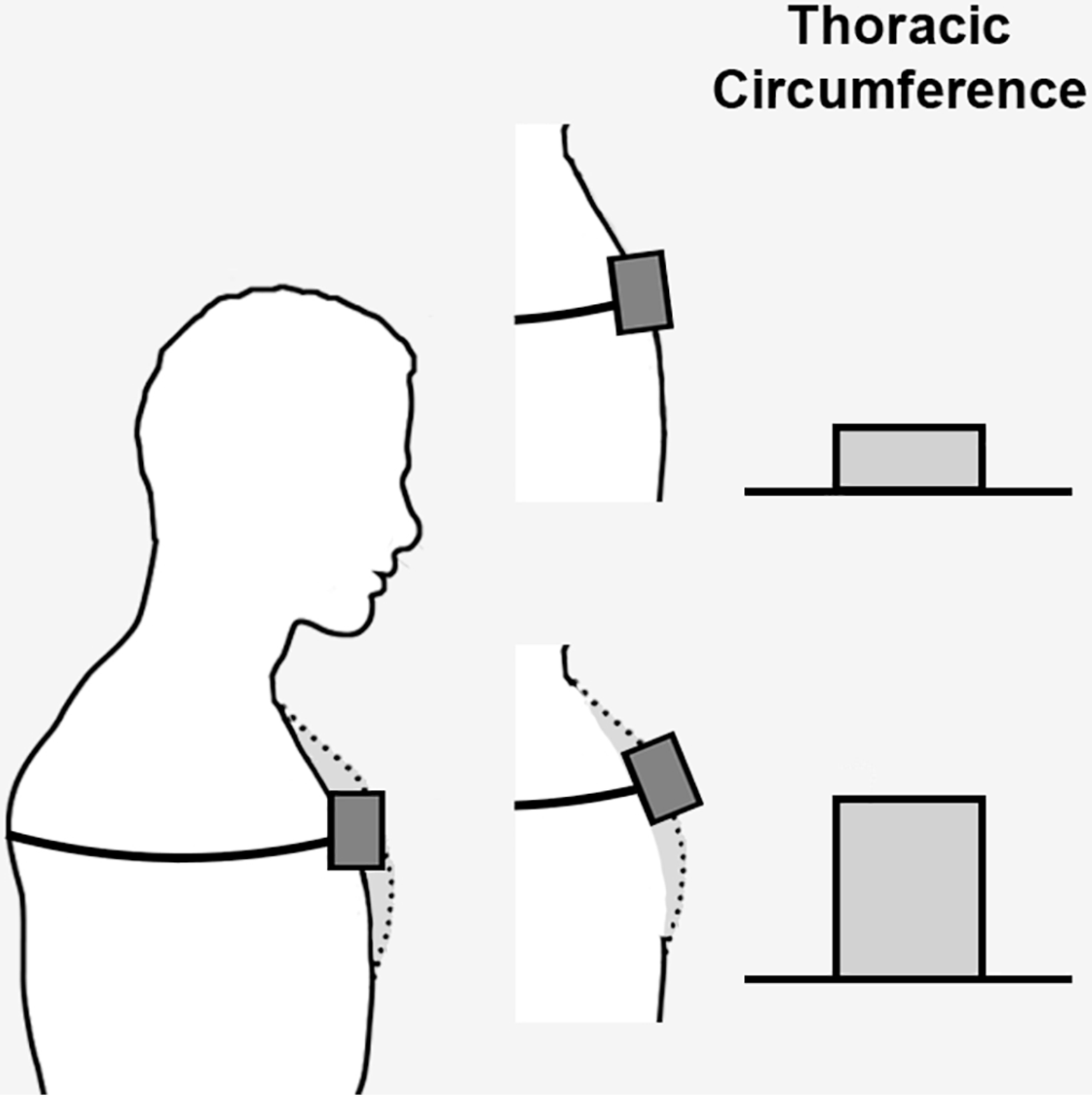}
\end{subfigure}
\hspace{\fill}
\begin{subfigure}[t]{0.23\textwidth}
\includegraphics[width=\textwidth]{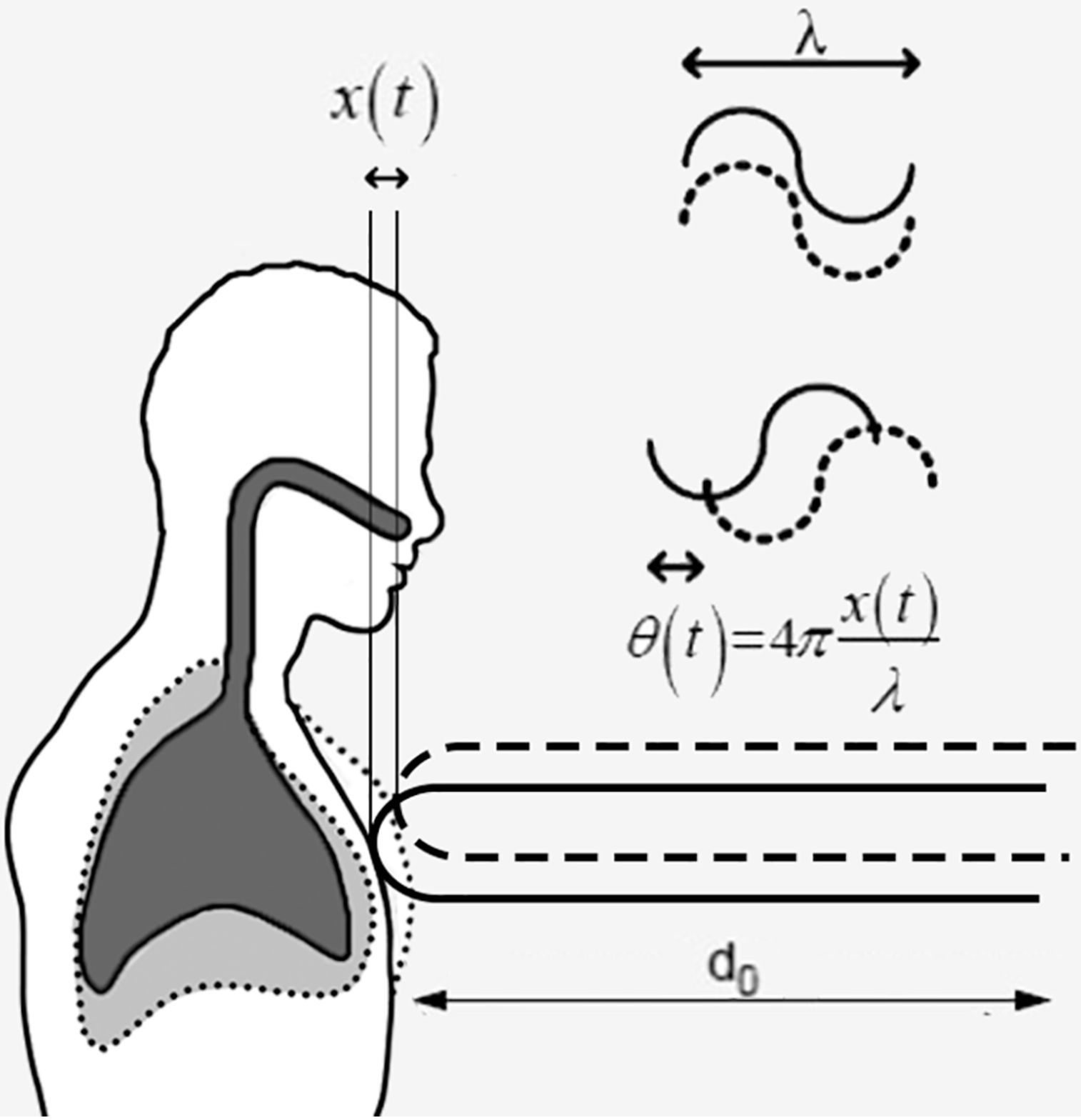}
\end{subfigure}
\caption{Left to right: (a) The respiratory belt detects chest displacement (breathing) through changes in thoracic or abdominal circumference; (b) The PRMS detects displacement (breathing) in patient's chest through phase offset between the TX and RX signal.}
\label{fig:preliminary}
\vspace{-15pt}
\end{figure}

Before introducing SIENNA, we briefly review the mechanisms of two common at-home OSA screening modalities: respiratory belt and non-contact PRMS. An external motion respiratory belt sensor utilizes a transducer to generate a substantial linear signal in response to changes in thoracic circumference associated with respiration (Fig. \ref{fig:preliminary}a). The linear signal is first sampled by a analog-to-digital converter (commonly at 100 Hz), then transmitted to a mobile OSA app.

The PRMS (Fig. \ref{fig:preliminary}b) utilizes Continuous Wave (CW) Doppler radar technology to detect the phase shift of reflected signals from the patient's chest movements. Let the distance offset due to chest movements be $x(t)$, and the in-phase (\textit{I}) and quadrature phase (\textit{Q}) can expressed as:
\begin{align*}
  {B}_I(t) &= A_I\cos\left(\theta_0+\frac{4\pi x(t)}{\lambda}+\delta\theta(t)\right)\\
  {B}_Q(t) &= A_Q\sin\left(\theta_0+\frac{4\pi x(t)}{\lambda}+\delta\theta(t)\right)
\end{align*}
where $\lambda$ is the signal wavelength, $\theta_0$ is the phase delay due to the nominal distance between the radar transmitter and the user's torso, surface scattering, and radar's RF chains, and $\delta\theta(t)$ is the residual phase noise. The phase shift corresponds to the respiratory movement and can be computed via arctangent demodulation:
\begin{equation*}
  \theta(t) = \theta_0 + \frac{4\pi x(t)}{\lambda} = \arctan \left(\frac{A_I{B}_Q(t)}{A_Q{B}_I(t)}\right).
\end{equation*}

\section{Problem Description}
\label{sec:challenges}

A respiratory belt and a PRMS need to pair with the user's mobile phone before an OSA test.  A respiratory belt is often paired with the user's phone by a medical technician during a clinic visit. The PRMS is usually shipped directly to the user's home and paired without supervision. The unsupervised pairing process is subject to exploitation from a non-compliant user. Assuming the respiratory belt has successfully paired with the phone,  we aim to enable automatic device paring between the PRMS and the phone via the shared context information, e.g., the user's respiratory patterns observed by the belt and PRMS. Unlike previous works on context-based zero-effort pairing, our pairing protocol must pair two devices securely in the presence of a co-located adversary who can also observe the context information.



\subsection{System Model}
We consider a multimodality OSA screening system with three modules: (1) A mobile phone that aggregates the screening data, a PRMS, and a wireless respiratory belt, and assume the following. (1) Wireless interface: The phone, PRMS, and belt are equipped with radio interfaces such as Bluetooth. (2) Computation: The PRMS and belt can perform computational inexpensive cryptographic algorithms, such as SHA-256 hash and AES. (3) Tamper-proof: The phone, PRMS, and belt are tamper-proof. Any attempts to physically modify the circuit would nullify the test. (4) Security: The phone, PRMS, and belt do not have any prior security associations. Secret keys are established between the belt and the user's phone by a medical technician (Fig. \ref{fig:sienna overview}a).

\subsection{Adversary Model}
A distinguishing feature of our adversary model is that the system's legitimate user could also be an insider attacker (non-compliant user). The attacker's objective is to either eavesdrop on the communication between the system modules or manipulate the system into accepting false data.

\textbf{Eavesdrop.} A non-compliant user may seek to eavesdrop on the pairing communication between the PRMS and the phone, aiming to extract the security context. For instance, the patient may intercept the key exchanged between the PRMS and the phone to decrypt and review all data records before a doctor examines. If patterns related to OSA symptoms were found, the patient would have time to come up with false excuses.

\textbf{Spoofing.} A non-compliant user may leverage the eavesdropped key to transmit false data to the mobile device and manipulate the OSA test outcome. For instance, the patient may use a third device to collect normal patterns prior to the test and replay the normal data records to the phone during the test.





\section{{SIENNA}}
\label{sec:sienna}

\begin{figure*}[t]
\begin{subfigure}[t]{0.24\textwidth}
\includegraphics[width=\textwidth]{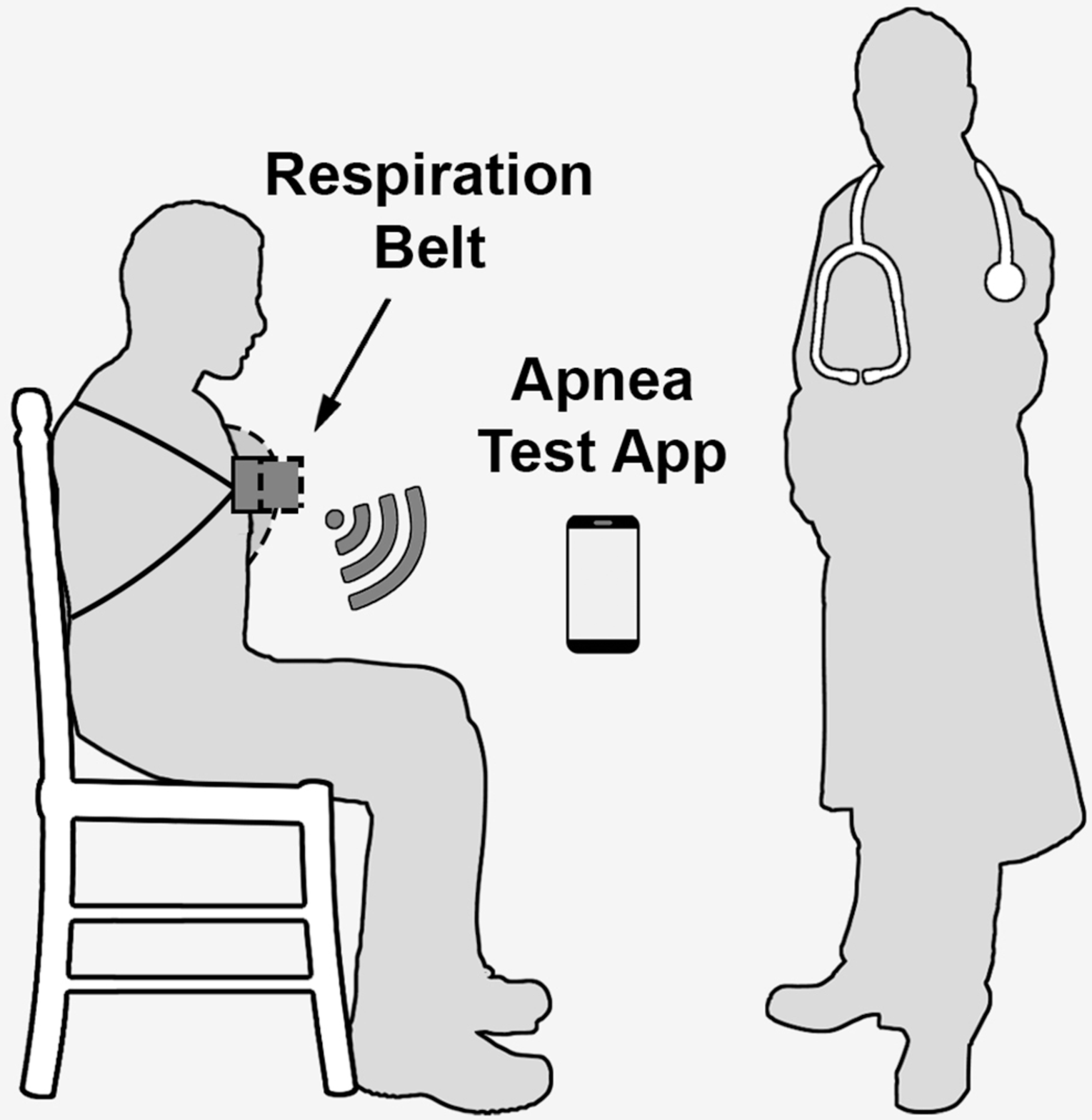}
\end{subfigure}
\hspace{\fill}
\begin{subfigure}[t]{0.24\textwidth}
\includegraphics[width=\textwidth]{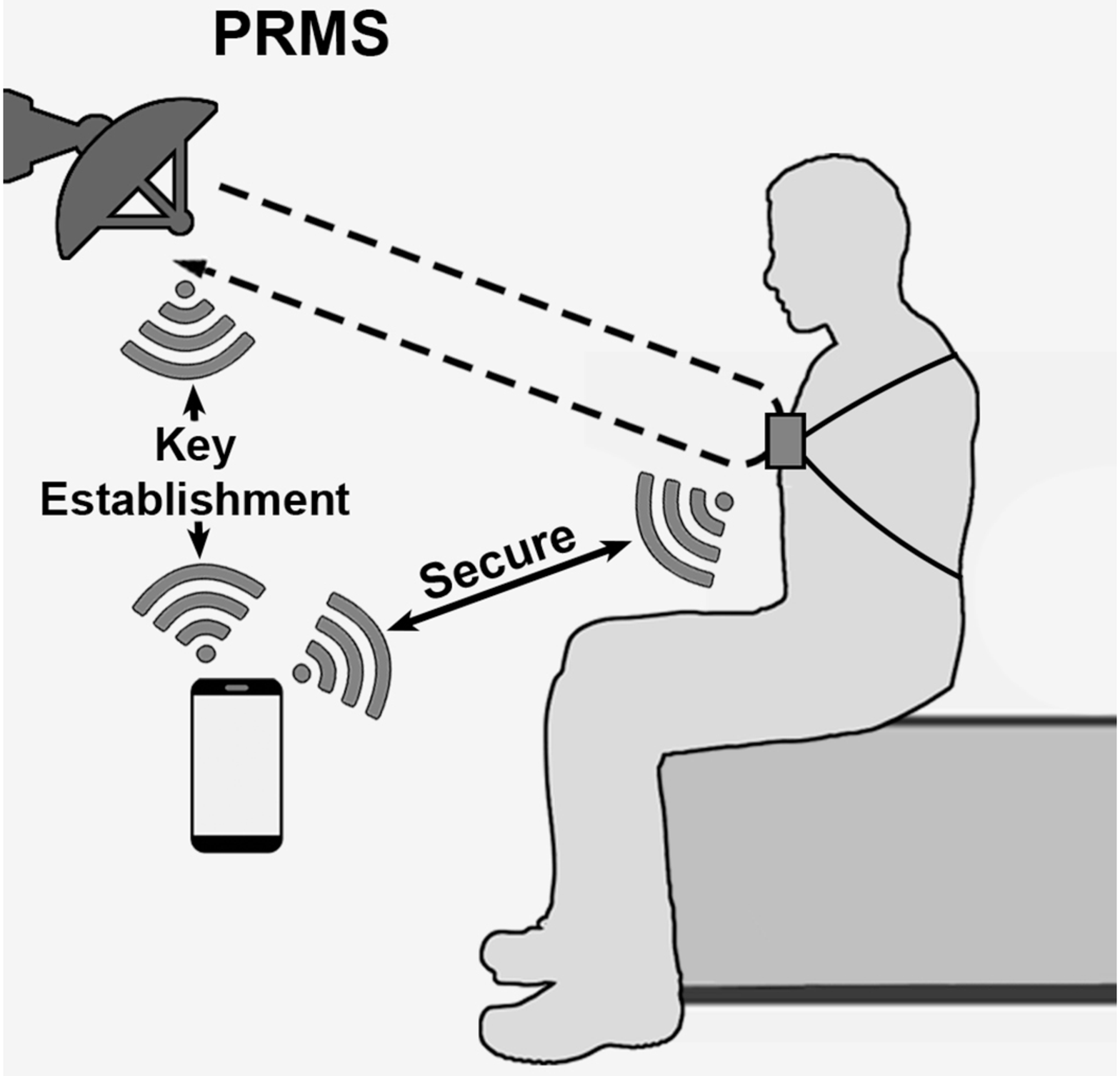}
\end{subfigure}
\hspace{\fill}
\begin{subfigure}[t]{0.24\textwidth}
\includegraphics[width=\textwidth]{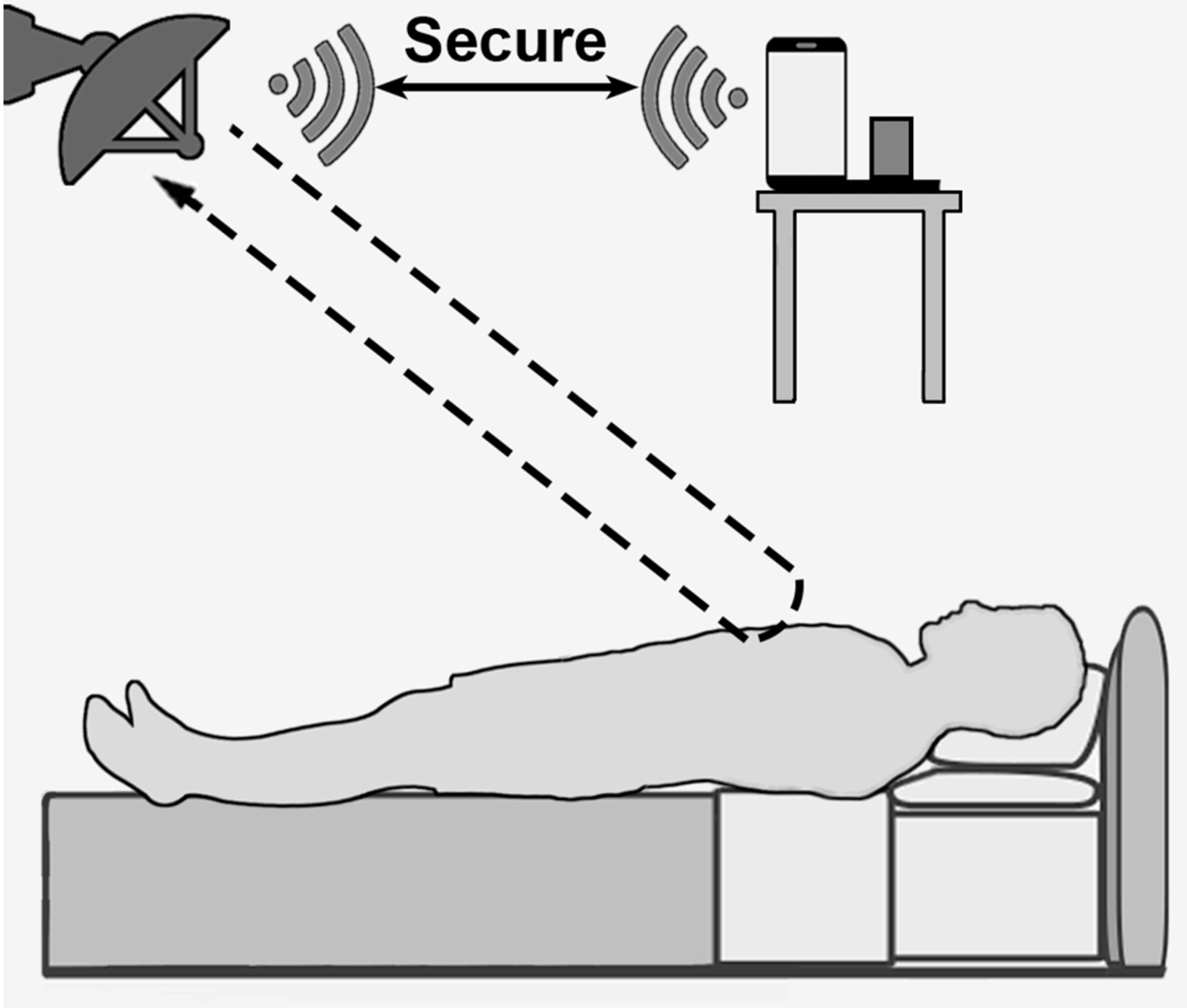}
\end{subfigure}
\hspace{\fill}
\begin{subfigure}[t]{0.24\textwidth}
\includegraphics[width=\textwidth]{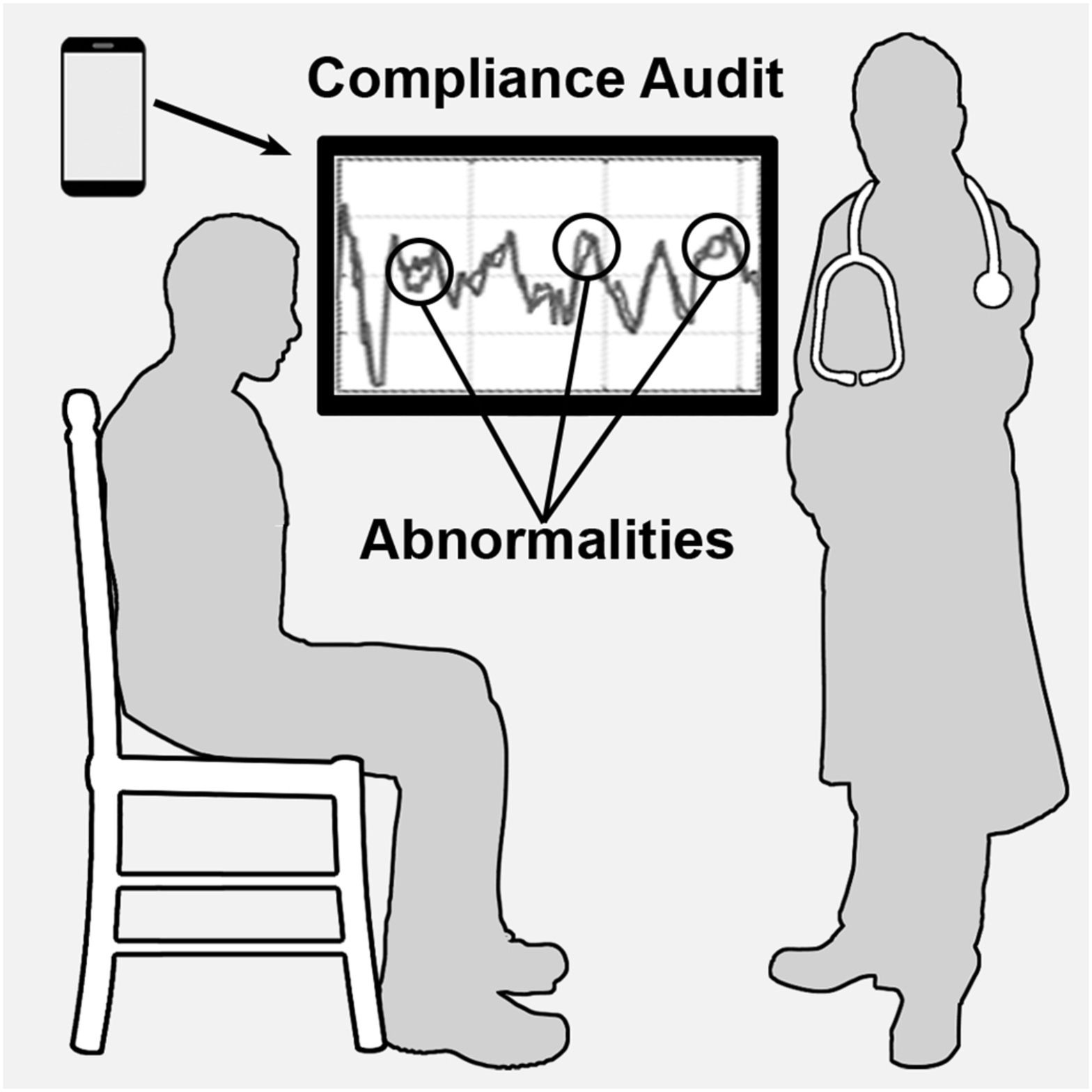}
\end{subfigure}
\caption{Left to right: (a) In hospital: a medical technician places a respiratory belt on a user and pairs it with the user's phone; (b) At home: a user pairs the PRMS with the user's phone; (c) User switches to PRMS for OSA screening; (e) Doctor verifies OSA data for any abnormalities.}
\label{fig:sienna overview}
\end{figure*}

We present SIENNA: in\textbf{SI}der r\textbf{E}sista\textbf{N}t co\textbf{N}text-based p\textbf{A}iring. SIENNA leverages the user's respiratory patterns observed by both the respiratory belt and the PRMS. To defense against an insider attack, SIENNA employs a combination of fuzzy matching and friendly jamming to prevent a non-compliant user from obtaining the pairing key using the same context.

\subsection{Overview}
The pairing procedure of SIENNA is shown in Fig. \ref{fig:sienna overview}. It begins when the user visits a doctor to obtain the test authorization. During the visit, the doctor attaches a respiratory belt to the patient, and pairs it to the user's mobile device OSA app (Fig. \ref{fig:sienna overview}a). Once arriving home, the user lies in bed and the PRMS automatically pairs with the mobile device based on the respiration pattern observed by both the PRMS and the respiratory belt (Fig. \ref{fig:sienna overview}b). Once the pairing completes, both links from the PRMS and respiration belt to the mobile device are secure. The user can freely choose either the respiratory belt or PRMS for OSA screening, and the selected modality communicates encrypted OSA data to the mobile device (Fig. \ref{fig:sienna overview}c). Once testing is completed, the user revisits the doctor and uploads the OSA screening from the mobile device. The doctor runs a compliance check and examines whether there were any significant gaps or inconsistencies with the OSA data. Based on the compliance check report, the doctor decides whether to accept or reject the OSA screening (Fig. \ref{fig:sienna overview}d).

\subsection{Insider Resistant Device Pairing}
\label{section:key evolution}
\begin{figure}[h]
\centering
\includegraphics[width=0.65\linewidth]{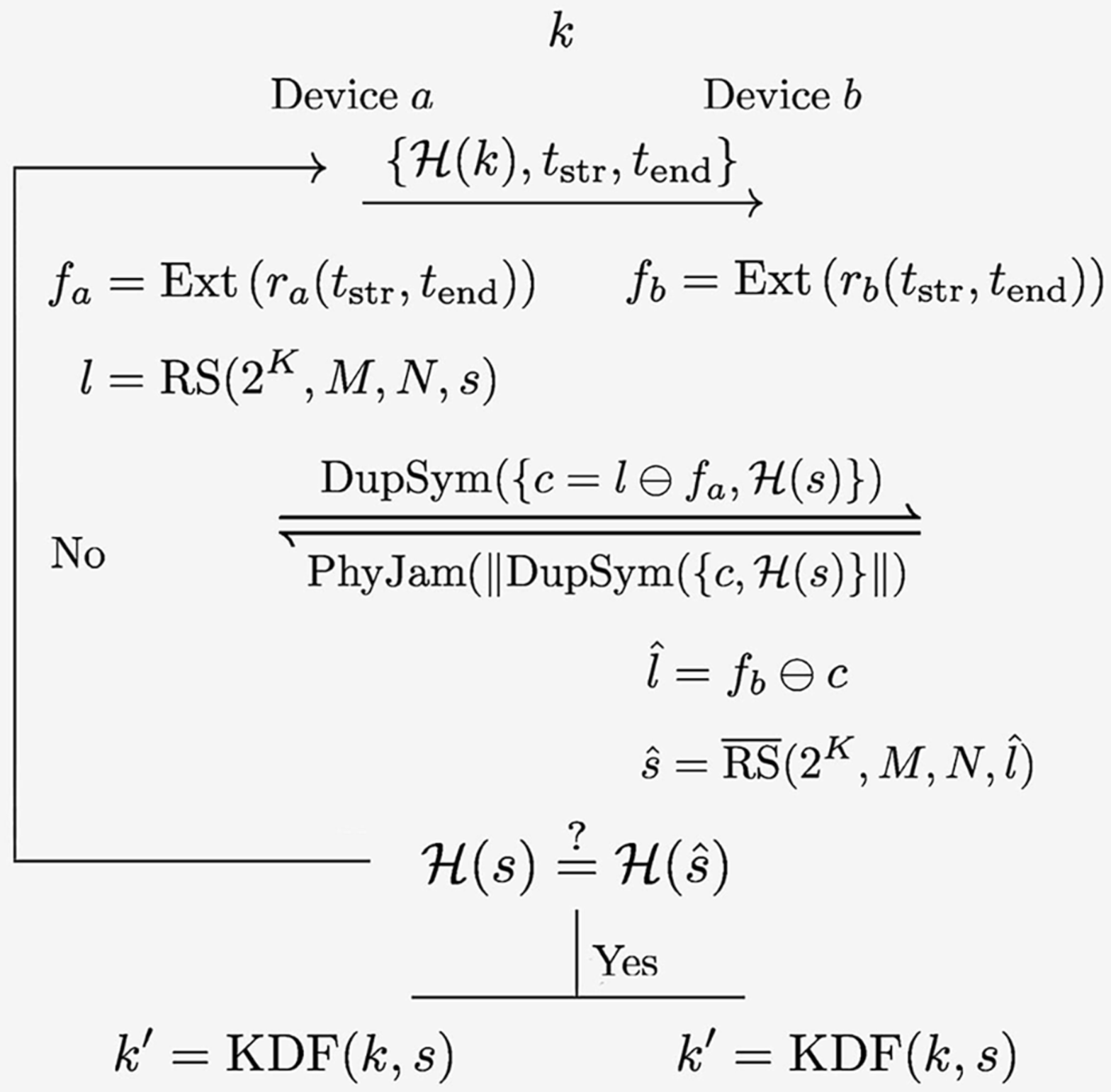}
\caption{Pairing protocol for two devices observing similar data to establish a symmetric key in the presence of an adversary.}
\label{fig:key evolution_diag}
\end{figure}

The core of SIENNA is a context-based secure key establishment protocol (Fig. \ref{fig:key evolution_diag}), which allows two devices, $a$ and $b$, to securely exchange and update a symmetric key in the presence of a nearby eavesdropper by utilizing the context (user's breathing patterns) observed by both devices at the moment of exchange. In the OSA scenario above, $a$ represents the mobile device connected to a respiratory belt, $b$ represents the PRMS.

Traditional context-based key establishment protocols are not secure when eavesdroppers are nearby to observe the context. Many of such protocols assume that the adversary can't be near paring devices over extended periods of time \cite{MiettinenContextBasedZeroInteractionPairing2014}. SIENNA addresses this shortcoming through a cross-layer design that employs two security primitives: fuzzy commitment \cite{JuelsFuzzyCommitmentScheme1999} and dialog-codes-based friendly jamming \cite{AroraDialogCodesSecure2009,GollakotaPhysicalLayerWireless2011, MelcherIJamChannelRandomization2020}. The former is a cryptographic scheme that allows secure commit and de-commit of a secrete similar-but-not-identical open value. The latter is a friendly jamming scheme at the receiver by jamming the transmitted signal to flip specific bits in the message. Together, this ensures that an eavesdropper with the right context will not be able to recover the key exchanged between the pairing devices.

First we define the fuzzy commitment scheme \cite{JuelsFuzzyCommitmentScheme1999}. Let $\sigma$ be a secret value. A fuzzy commitment transforms a secret value $\sigma$ into a commitment $\{\chi, \mathcal{H}(\sigma)\}$ using an opening feature, $\phi$, and a hash function, $\mathcal{H}(\cdot)$:
\[
\{\chi, \mathcal{H}(\sigma)\} = \textsc{Commit}(\sigma, \phi),
\]
such that $\chi$ appears random and devoid of any information about $\sigma$. And all open features $\hat{\phi}$ reveals $\sigma$ via 
\[
\sigma=\textsc{Open}(\chi,\hat{\phi}),
\]
if and only if the Hamming distance $\textsc{Ham}(\phi, \hat{\phi}) \leq \tau$, where $\tau$ is a parameter denoting the maximum allowable Hamming distance between $\hat{\phi}$ and $\phi$ to reveal $\sigma$. 

To initiate the key establishment protocol, $a$ broadcasts a message, $\{ \mathcal{H}(k), t_{\text{str}}, t_{\text{end}} \}$, where $k$ denotes the key of the previous iteration. $t_{\text{str}}$ and $t_{\text{end}}$ denotes the starting and ending timestamps of $a$ and $b$'s captures of the respiratory patterns. During the first round,  $\mathcal{H}(k)$ is replaced with a public parameter known to all parties.

From respiratory patterns $r_{a}(t_{\text{str}}, t_{\text{end}})$ and $r_{b}(t_{\text{str}}, t_{\text{end}})$ captured within the specified time interval, the two devices extract breathing fingerprints $f_{a} = \textsc{Ext} \left( r_{a}(t_{\text{str}}, t_{\text{end}}) \right)$ and $f_{b} = \textsc{Ext} \left( r_{b}(t_{\text{str}}, t_{\text{end}}) \right)$, via a fingerprint extraction function $\textsc{Ext} (\cdot): \{1,0\}^{*} \mapsto \{1,0\}^{M*2^K}$ (detailed in Sec. \ref{section:respiration fingerprinting}). In case $a$, e.g., PRMS, observes a breathing mixture of multiple subjects, the mixture is first separated into the breathing patterns of individual subjects (detailed in Sec. \ref{section:breathing separation}), which are processed by $\textsc{Ext} (\cdot)$ to create multiple fingerprints, one for each subject.

Once the breathing fingerprints are generated, $a$ randomly selects a key salt, $ s \in \{1,0\}^{N*2^K}$, and transforms it into a commitment $\{c \in \{1,0\}^{M*2^K}, \mathcal{H}(s)\}$ using the breathing fingerprint, $f_{a}$. Specifically, $a$ encodes $s$ via the Reed-Solomon (RS) encoding function
\[
l = \textsc{RS}(2^K, M, N, s) \in \mathbb{F}^M_{2^K}
\]
and compute:
\[
c = l \ominus f_{a},
\]
with $\ominus$ denoting exclusive OR (XOR).

Henceforth, $a$ and $b$ exchange $c$ through dialog codes to defend against insider attack. First, $a$ converts the commitment,
\[
\{c, \mathcal{H}(s)\},
\]
into OFDM symbols, duplicates each symbol back-to-back,
\[
\textsc{DupSym}\left(\{c, \mathcal{H}(s)\}\right),
\]
and broadcasts all the symbols. In parallel to $a$'s broadcast, $b$ randomly jams either the original symbol or its repetition \cite{GollakotaPhysicalLayerWireless2011,AroraDialogCodesSecure2009}, 
\[
\textsc{PhyJam}\left(\|\textsc{DupSym}\left(\{c, \mathcal{H}(s)\}\right)\|\right),
\]
To jam a symbol, $b$ transmits a signal that is drawn randomly from a zero-mean Gaussian distribution whose variance is the same as the OFDM signal with the same modulation. Since $b$ knows which symbols are jammed, it stitches the unjammed symbols together to create a clean version of the OFDM transmission and decodes the signal to obtain the clear message.

Upon receiving $\{c, \mathcal{H}(s)\}$, $b$ computes 
\[
\hat{l} = f_{b} \ominus c
\]
and decommits the salt by decoding $\hat{l}$ using the Reed-Solomon (RS) decoding function: 
\[
\hat{s} = \overline{RS}(2^K, M, N, \hat{l}).
\]
Due to the error correction capability of Reed–Solomon codes, $s$ equals to $ \hat{s}$ if and only if $l$ and $\hat{l}$ differ in less than $2^{K-1}(M-N)$ bits. Since $l = f_{a} \ominus d$ and $\hat{l} = f_{b} \ominus d$, it is equivalent in saying that $f_{a}$ and $f_{b}$ must differ in less than $T$ bits for $b$ to retrieve $s$.

To confirm whether the retrieval were successful, $b$ computes $\mathcal{H}(\hat{s})$ and compares it with $\mathcal{H}(s)$. Depending on whether they were equal, an $\textsc{ACK}$ or a $\textsc{NAK}$ message is transmitted from $b$ to $a$, with the former initiating the final step of the key establishment protocol and the latter initiating the reattempts.

To conclude the key establishment, both $a$ and $b$ applies a key derivation function 
\[
k' = \textsc{KDF}(k, s),
\]
to obtain the new key.

\subsection{Breathing Separation with JADE-ICA}
\label{section:breathing separation}
Home environments are noisy and unpredictable, with the possibility of irrelevant individuals in close vicinity from the user. To retrieve the correct context in an environment with potentially multiple subjects, SIENNA augments the PRMS modality with a breathing separation module, which reconstructs the breathing signals of multiple co-located individuals to select the correct target. 
The goal of the separation module is to reconstruct a set of source signals from a set of mixtures, without knowing the properties of the sources and the mixing proportion. 
Since respiration signals are non-Gaussian and independent from individuals, whilst mixed linearly at the PRMS receiver, one can recover the source signals using independent component analysis (ICA) \cite{yue_extracting_2018}, which is formulated as the following: assume $N$ independent time varying sources $ s_i(t),\ i=1 \ldots N$, and $M$ different observations
$ x_i(t),\ i=1 \ldots M$. For $T$ time units $(t=1 \ldots T)$, we can define the source signal as a $N \times\ T$ matrix,
\[
\mathbf{S}_{N \times T} = \begin{bmatrix} 
    s_{11} & s_{12} & \dots \\
    \vdots & \ddots & \\
    s_{N_1} &        & s_{N_T} 
    \end{bmatrix},
\]
and the observed mixtures as a $M \times\ T$ matrix,
\[
\mathbf{X}_{M \times T} = \begin{bmatrix} 
    x_{11} & x_{12} & \dots \\
    \vdots & \ddots & \\
    x_{M_1} &        & x_{M_T}
    \end{bmatrix}.
\]
The mixtures are produced as the product of the source and a mixing matrix $\mathbf{W}_{M \times N}$, e.g.,
\[
    \mathbf{X}_{M \times T} = \mathbf{W}_{M \times N} \times \mathbf{S}_{N \times T}.
\]
The goal of ICA is to recover $\mathbf{S}_{N \times T}$ and $\mathbf{W}_{M \times N}$ given only $\mathbf{X}_{M \times T}$, assuming
the $s_i(t),\ i=1 \ldots N$ are independent and non-Gaussian. We employ the joint approximate diagonalization of eignematrices (JADE) algorithm \cite{de_lathauwer_independent_1996} to perform ICA, with the details omitted to conserve space. 

The JADE algorithm extracts independent non-Gaussian sources $\mathbf{S}_{N \times T}$, e.g., the breathing pattern of individual targets, from signal mixtures with Gaussian noise $\mathbf{X}_{M \times T}$, e.g., the breathing mixture received by the PRMS, by constructing a fourth-order cumulants array from $\mathbf{X}_{M \times T}$. Specifically, assuming $T > M,N$, the algorithm first applies PCA to whiten $\mathbf{X}_{M \times T}$, resulting
\[
\mathbf{P}^{\text{tr}}_{T \times K} = \mathbf{B}_{K \times M} \times \mathbf{X}_{M \times T},
\]
with a whitening matrix $\mathbf{B}_{K \times M}$.
After the whitening, the columns of whitened matrix $\mathbf{P}_{T \times K}$ are orthogonal with equal variance. Any rotation of $\mathbf{P}_{T \times K}$ will not change the independence between it column vectors. The algorithm then tries to find a rotation matrix $\mathbf{V}_{N \times N}$ to maximize the independence between the row vectors of the rotated $\mathbf{P}_{T \times K}$, which is achieved when the fourth-order cross-cumulants between the row vectors are zero and their auto-cumulants maximal. Once $\mathbf{V}_{N \times N}$ has been identified, the demixing matrix can be computed as 
\[
    \mathbf{W}^{+}_{N \times M} = \mathbf{V}_{N \times N} \times \mathbf{B}_{N \times M},
\]
and the source matrix as
\[
    \mathbf{S}_{N \times T} = \mathbf{W}^{+}_{N \times M} \times \mathbf{X}_{M \times T},
\]
More detailed information on JADE and cumulant tensor array can be found at \cite{rutledge_independent_2013} and the Matlab code on the website \cite{CardosoBlindSeparationReal}.
 
 
 

\subsection{Fingerprinting with Level-Crossing Quantization}
\label{section:respiration fingerprinting}


Once the devices obtain breathing patterns of individual targets, they apply $\textsc{Ext} (\cdot)$ to extract the binary breathing fingerprints. The binary strings must meet two criteria for $a$ and $b$ to agree on the patient's identity and evolve the shared security key: (1) they should look sufficiently similar in Hamming space if they represent the breathing process of the same person, and (2) they should preserve the uniqueness of the breathing dynamic that distinguishes among individuals.

\begin{figure}[h]
\centering
\includegraphics[width=0.65\linewidth]{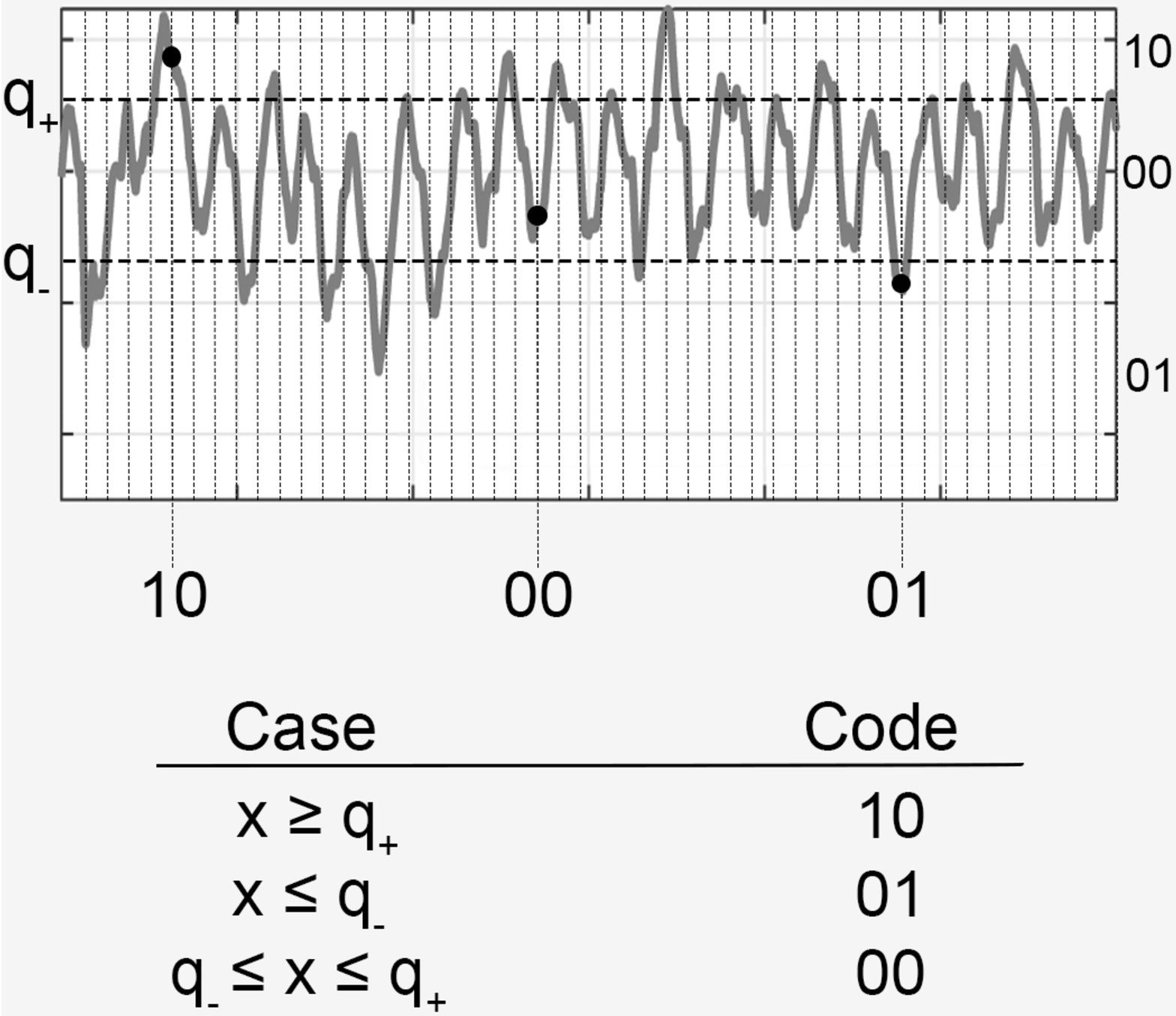}
\caption{Encoding of a breathing pattern signal \& generation of a key using level-crossing quantization.}
\label{fig:quantization}
\end{figure}

To achieve these objectives, $\textsc{Ext} (\cdot)$ applies level-crossing quantization to sample the continuous breathing patterns with two predefined thresholds (Fig. \ref{fig:quantization}). Let $q_{+}$, $q_{-}$ be the thresholds values such that $q_{+} > q_{-}$, we define a quantizer $\textsc{Qtz} (\cdot)$:
\begin{equation*}
  \setlength{\arraycolsep}{0pt}
    \textsc{Qtz} (x) = \left\{ \begin{array}{ l l }
    10 &\;\;\;\;\text{if}\; x \geq q_{+} \\
    01 &\;\;\;\;\text{if}\; x \leq q_{-} \\
    00 &\;\;\;\;\text{if}\; q_{-} < x < q_{+}
  \end{array} \right.
\end{equation*}
Let $T$ be the time interval between adjacent sampling instants. The binary sequence obtained by $\textsc{Ext} (\cdot)$ is
\begin{multline*}
f = \textsc{Ext} \left(r(t_{\text{str}}, t_{\text{end}})\right) = [\textsc{Qtz} \left( r(t_{\text{str}}) \right),...,\\
    \;\;\;\;\;\;\;\textsc{Qtz} \left(r(t_{\text{str}} + \lfloor\frac{t_{\text{end}} - t_{\text{str}}}{T}\rfloor T \right)],
\end{multline*}
which can be compared in Hamming space.
If the binary sequence is longer than the length of the commitment,  $\|\lambda\|$, we pad and divide it into multiple subsequences, $f = [ f_1, f_2,...,f_n]$ to commit
\begin{equation}
c = f_1 \ominus f_2 \ominus ... \ominus f_n \ominus l,
\end{equation}
and decommit
\begin{equation}
\hat{l} = \hat{f}_1 \ominus \hat{f}_2 \ominus ... \ominus \hat{f}_n \ominus c,
\end{equation}
The level-crossing quantization described above is an event-based sampling procedure as the crossing events are encoded as changes in the binary patterns. When the signal remains above/below the levels, it produces fixed patterns, which can be exploited to further compress the binary before transmission\footnote{Techniques to produce a compressed the binary string that are satisfiable to the two criteria is beyond the scope of this article and will be investigated in the future.}.

However, the result of a single level-crossing quantization loses details in the original breath pattern and fails the second objective. To address this issue, we apply multiple passes of level-crossing binary quatizations, each at a distinct pair of levels, $q_{i+}$, $q_{i-}$. Intuitively, it is equivalent to create a pair-wise linear approximation of the original breathing pattern, with quantization error equal to the level density.

If the binary fingerprints after the multi-level quantization is longer than $\|l\|$, we pad and divide it into multiple subsequences, $f = [ f_1, f_2,...,f_n]$ to commit
\[
c = f_1 \ominus f_2 \ominus ... \ominus f_n \ominus l,
\]
and decommit
\[
l = f_1 \ominus f_2 \ominus ... \ominus f_n \ominus c.
\]

\section{Security Analysis}
\label{sec:security_analysis}

The security of SIENNA can be formally analyzed based on the property of a fuzzy commitment, and extended into three cases according to the attacker's knowledge on the user's breathing fingerprint, $f$.

\subsection{Concealment and Binding}
Following the analysis conventions of the fuzzy commitment scheme by Juels and Wattenberg \cite{JuelsFuzzyCommitmentScheme1999}, we employ two metrics to analyze SIENNA's key evolution protocol, e.g., \textit{concealment} and \textit{binding}. 

\begin{definition}
\label{def:concealment and binding}
A binary fuzzy commitment, $\{\chi, \mathcal{H}(\sigma)\} = \textsc{Commit}(\sigma, \phi)$ where $\sigma \in_{R} \{1,0\}^{\nu}$ and $\mathcal{H}(\cdot) : \{1,0\}^{*} \mapsto \{1,0\}^{\mu}$,
is said to be concealing if it is infeasible for any polynomially-bounded player to guess $\sigma$ with probability greater than $p = 1/2^{\nu}$. It is said to be binding if it is infeasible for any polynomially-bounded player to retrieve $\sigma$ using a incorrect opening feature, e.g., $\hat{\phi},\; \textsc{Ham}(\phi, \hat{\phi}) > \tau$, with probability greater than $p = 1/2^{\mu}$. 
\end{definition}

\subsection{Attacker without Knowledge of $f$}
When the eavesdropper does not have the correct context, SIENNA inherits the security properties of a fuzzy commitment, which we reiterate below, with the proof omitted to conserve space.
\begin{claim}
\label{clm:concealment}
Suppose $s \in_{R} \{1,0\}^{N*{2^K}}$ and $f \in_{R} \{1,0\}^{M*{2^K}}$, the fuzzy commitment construction used by SIENNA, with $l = \textsc{RS}(2^K, M, N, s) \in_{R} \mathbb{F}^M_{2^K}$ and $c = f \ominus l$, is concealing with $p = 1/2^{N*2^K}$ against an attacker with no prior knowledge of $f$.
\end{claim}



\begin{claim}
\label{clm:binding}
Suppose $s \in_{R} \{1,0\}^{N*{2^K}}$ and $f \in_{R} \{1,0\}^{M*{2^K}}$, the fuzzy commitment construction used by SIENNA, with $l = \textsc{RS}(2^K, M, N, s) \in_{R} \mathbb{F}^M_{2^K}$ and $c = f \ominus l$, is binding with $p = 1/2^{M*2^K}$ and $\tau = 2^{K-1}(M-N)$.
\end{claim}


Overall, Claim \ref{clm:concealment} and \ref{clm:binding} characterize the hardness for an attacker without prior knowledge of $f$ to determine $s$ from $\{c, \mathcal{H}(s)\}$, and identify two security parameters, $\nu$ and $\mu$, which govern the security level for concealment and binding. Assuming that the most effective means of finding a collision for a hash function is a birthday attack, which induces a work factor of $2^{\mu/2}$, we can set $\nu = 128$, and $\mu = 256$ to guarantee strong concealment and binding properties, with the hardness similar to finding a collision in SHA-256.

\subsection{Attacker with General Knowledge of $f$}
While the binding level of Claim \ref{clm:binding} holds regardless of the opening feature $f$'s probability distribution, the concealment level of Claim \ref{clm:concealment} would fall if $f$ is drawn from a non-uniform distribution known to the attacker. Specifically, by knowing the distribution of $f$, the attacker's strategy to determine $\hat{l}$ in the proof for Claim \ref{clm:concealment} can be computationally less expensive than inverting $\mathcal{H}(r)$ for a uniform random value, $r$. 



\subsection{Attacker with Perfect Knowledge of $f$}
The XOR-chain trick would not prevent an attacker with perfect knowledge of $f$ to retrieve $s$. Specifically, consider a malicious patient capable of measuring his own breathing patterns. Were he also able to capture the commitment message, $\{c, \mathcal{H}(s)\}$, he can accurately compute $l = c  \ominus f$ and decode to obtain $s$. To prevent such an insider attack, SIENNA leverages friendly friendly jamming at the physical layer to obfuscate the commit message for any unintended receivers.

While SIENNA's friendly jamming technique is universally applicable at the physical layer of any digital communication system, it is particularly effective when augmenting an OFDM system, due to pseudorandom character of the signal. During OFDM modulation, a binary sequence is converted into $N$ complex numbersin the frequency-domain, $X_n$, via quadrature amplitude modulation (QAM), then converted into a time-domain sequence, $x_k$ via the inverse fast Fourier transform (IFFT),
\[
x_k = \sum_{n=0}^{N} X_n e^{i2\pi k n / N}.
\]
Each $x_k$ can be regarded as a weighted sum of $N$ pseudorandom variables due to the IFFT, resulting a pseudorandom Gaussian signal according to the central limit theorem (CLT). When the jamming signal is drawn randomly from a zero-mean Gaussian with the same variance of the OFDM signal, a single-antenna attacker cannot distinguish the jammed and clear signal\footnote{The authors also have shown that channel-randomized version of the jamming scheme is robust against eavesdropping from a multi-antenna attacker \cite{SteinmetzerLockpickingPhysicalLayer2015, PanROBinKnownPlaintextAttack2020, MelcherIJamChannelRandomization2020}}, therefore cannot properly reconstruct $\{c, \mathcal{H}(s)\}$.

We can analyze the jamming protection against an insider attack based on a wiretap channel model \cite{LiangCompoundWiretapChannels2009}. Consider a non-compliant patient using an unauthorized receiver to intercept the commit message. We denote the main channel as the wireless channel between the $a$ and $b$ and the wiretap channel as the one between the unauthorized receiver and one of $a$ and $b$. The frequency-domain representation of the main channel is
\[
Y_{\text{main}} = X + \frac{P_0}{P_1}\mathcal{N}(0, \sigma^2_0),
\]
and the frequency-domain representation of the wiretap channel is
\[
Y_{\text{tap}} = X + \frac{P}{P_2}\mathcal{N}(0, \sigma^2),
\]
where $P_0$ and $\sigma^2_0$ denote the average power and variance of the intrinsic wireless noise, $P_1$ and $P_2$ denote the average powers of the OFDM signal observed by the receiver and the unauthorized receiver, and $P$ and $\sigma^2$ denote average power and variance of the jamming signal observed by the unauthorized receiver. The secrecy capacity \cite{WynerWiretapChannel1975} of the wiretap model is
\[
C_s = \left[\log\left(1 + \frac{P_1}{P_0}\right) - \log\left(1 + \frac{P_2}{P}\right)\right]^{+}.
\]

It has been shown in \cite{GollakotaPhysicalLayerWireless2011}, the jamming scheme works at its optimal when the OFDM system operates with high order modulation (at least QPSK), and $1 < P/P_2 < 9$.\footnote{The jamming signal is too weak to degrade the OFDM signal when $P/P_2 \leq 1$, and too strong to be indistinguishable from the OFDM signal when  $P/P_2 \ge 9$.} Therefore, SIENNA prohibits the transmit in BPSK at any SNR. The bit error probability for such an OFDM system, allowing only M-QAM transmission, is 
\[
B_{\text{main}} \simeq \frac{4}{\log_2{M}}\textsc{Q}\left( \sqrt{\frac{3 P_1 \log_2{M}}{P_0(M-1)}} \right)
\]
for the main channel and 
\[
B_{\text{tap}} \simeq \frac{4}{\log_2{M}}\textsc{Q}\left( \sqrt{\frac{3 P_2 \log_2{M}}{P(M-1)}} \right),
\]
for the wiretap channel, where $\textsc{Q}(\cdot)$ denotes the the tail distribution function of the standard normal distribution. The receiver may adjust $P$ to elevate $B_{\text{tap}}$ beyond the error correction capability of the fuzzy commitment, and prevent an insider attack. The issue is that the jamming is only effective when $1 < P/P_2 < 9$, but $P_2$ depends on the location of the unauthorized receiver and is unknown to the receiver. Our solution is to have the transmitter create $L$ commitments, each with one sub-salt, and transmit them one by one, while the receiver jams at $L$ different power levels, $\{P_{\text{max}}, P_{\text{max}}/9, \ldots, P_{\text{max}}/9^{L-1}\}$. Given the fact that $B_{\text{main}}$ is not affected by $P$, the receiver can recover all sub-salts and XOR them together to obtain the key evolution salt. In contrast, the unauthorized receiver will fail to decode at least one sub-salt, therefore cannot recover the key evolution salt. The number of jamming levels, $L$, can be computed based on the upper bound (the maximum power supported by the hardware, $P_{\text{max}}$) and lower bound (the noise floor, $P_0$) on the OFDM signal power.  

\section{Evaluation}
\label{sec:evaluation}
\begin{figure}[h]
\begin{subfigure}[t]{0.23\textwidth}
\includegraphics[width=\linewidth]{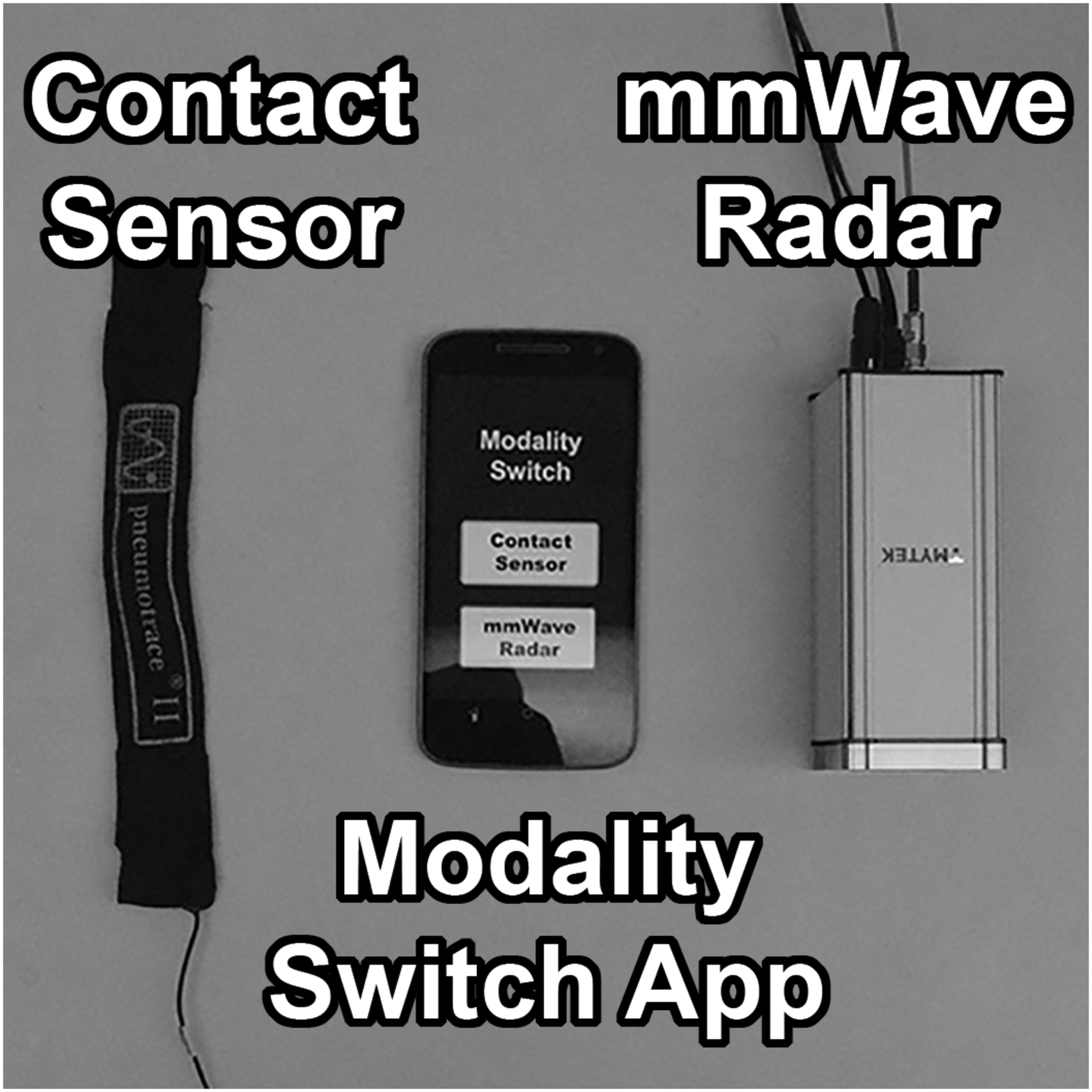}
\end{subfigure}
\hspace{\fill}
\begin{subfigure}[t]{0.23\textwidth}
\includegraphics[width=\textwidth]{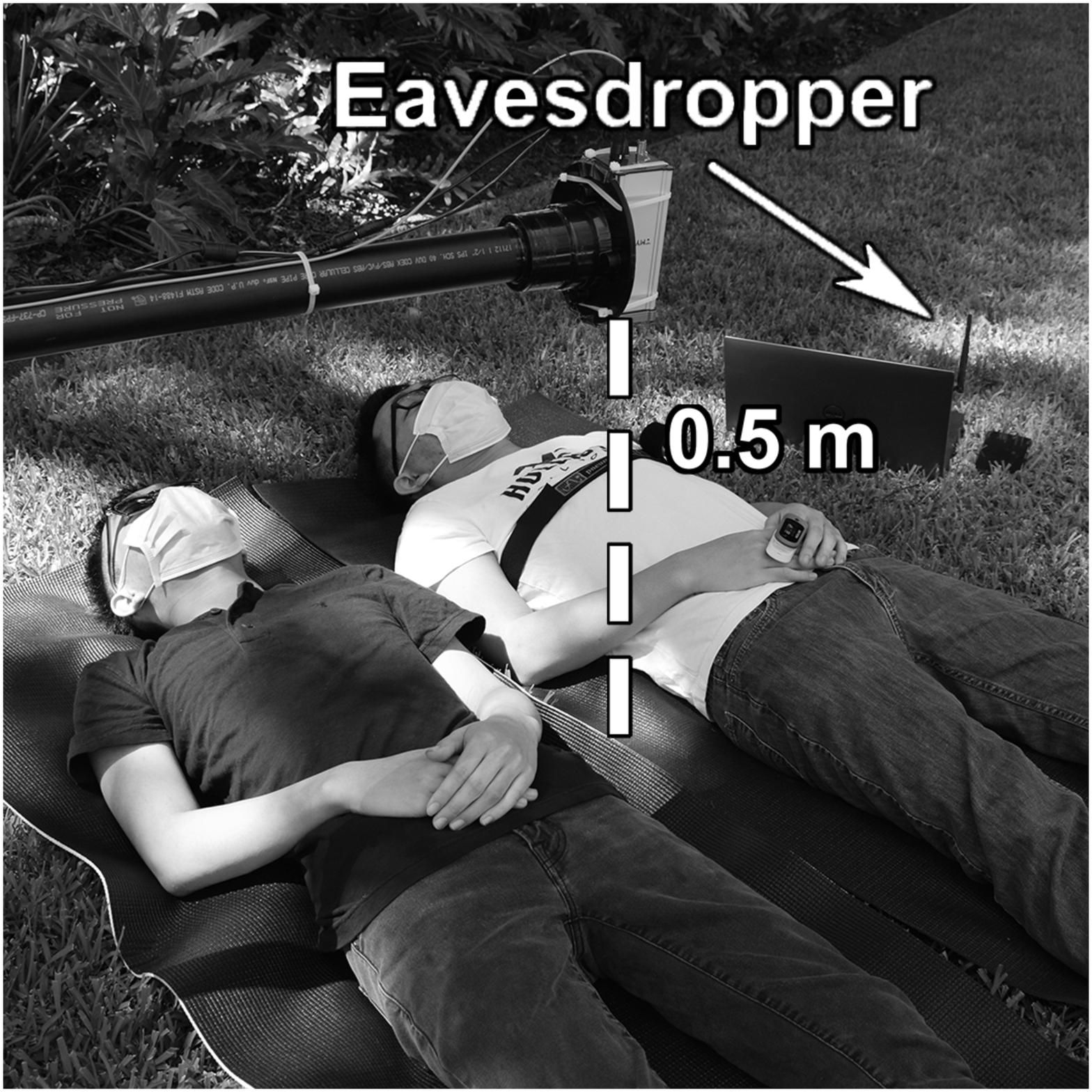}
\end{subfigure}
\caption{Left to right: (a) Implementation of SIENNA with, an Android application, a contact sensor, and PRMS; (b) Eavesdropper records signal from PRMS.}
\label{fig:experiment}
\end{figure}

We empirically evaluated the performance of SIENNA, which consists of a PRMS implemented with mmWave transceivers/radio heads, one respiration belt sensor implemented with a piezo-electric respiration transducer, and one Android-based OSA application (Fig. \ref{fig:experiment}a). We conducted laboratory and field experiments over one month with the SIENNA prototype and 20 subjects selected through a random sample recruitment process. All experiments with human subjects are approved by the Institutional Review Board (IRB) based on the written consent.

\begin{figure*}[t]
\begin{subfigure}[t]{0.24\textwidth}
\includegraphics[width=\textwidth,height=\textwidth]{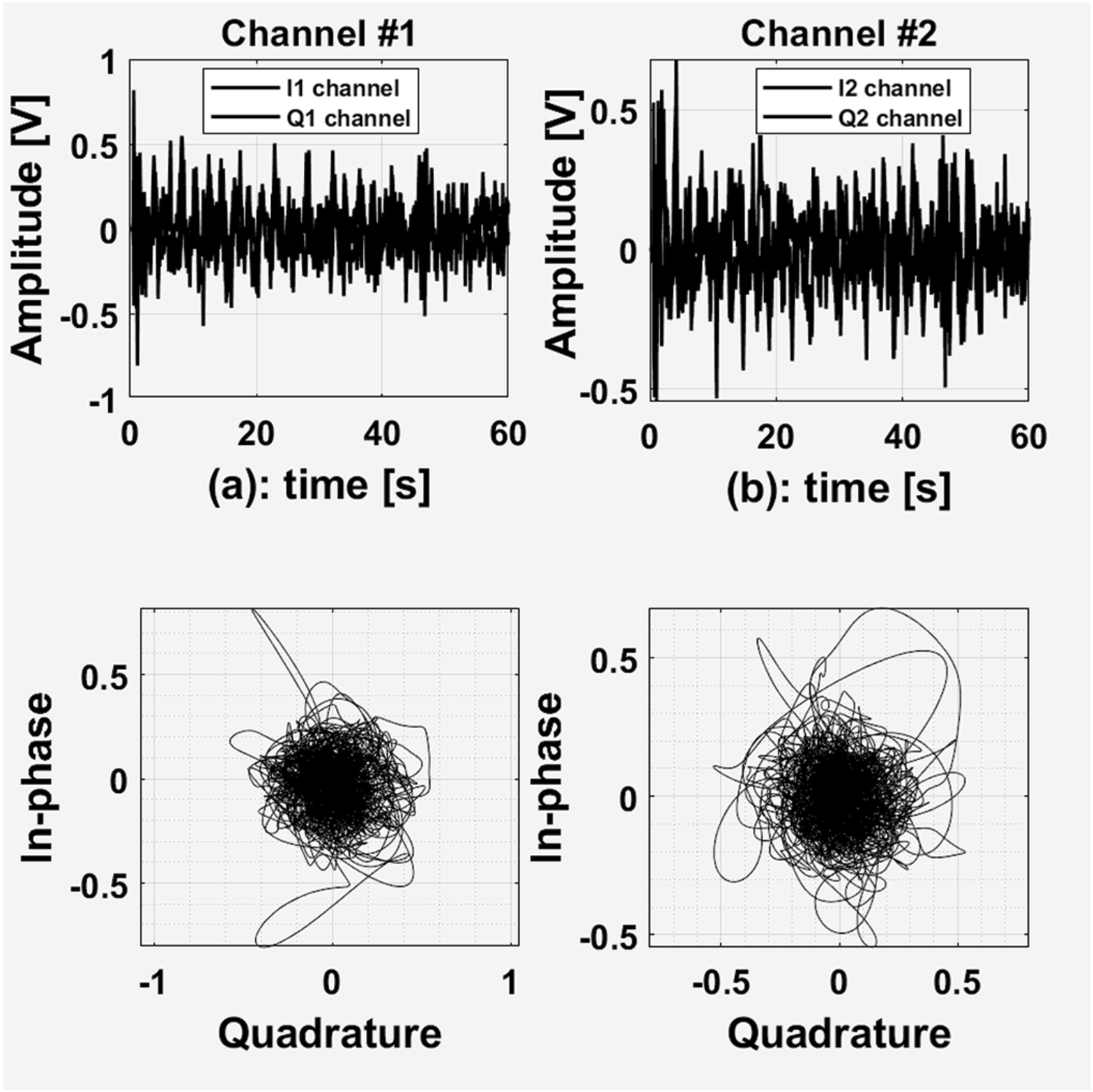}
\end{subfigure}
\hspace{\fill}
\begin{subfigure}[t]{0.24\textwidth}
\includegraphics[width=\textwidth,height=\textwidth]{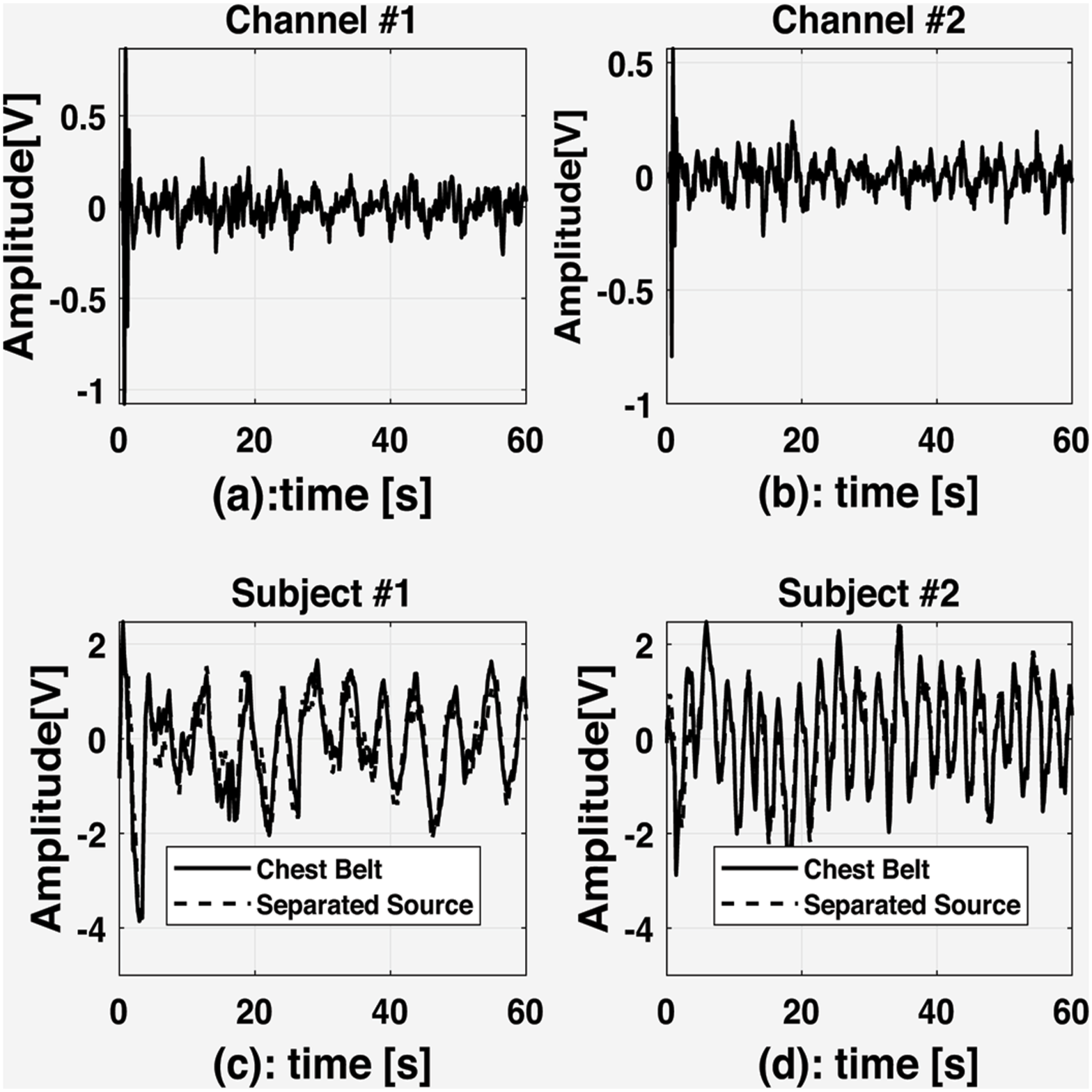}
\end{subfigure}
\hspace{\fill}
\begin{subfigure}[t]{0.24\textwidth}
\includegraphics[width=\textwidth,height=\textwidth]{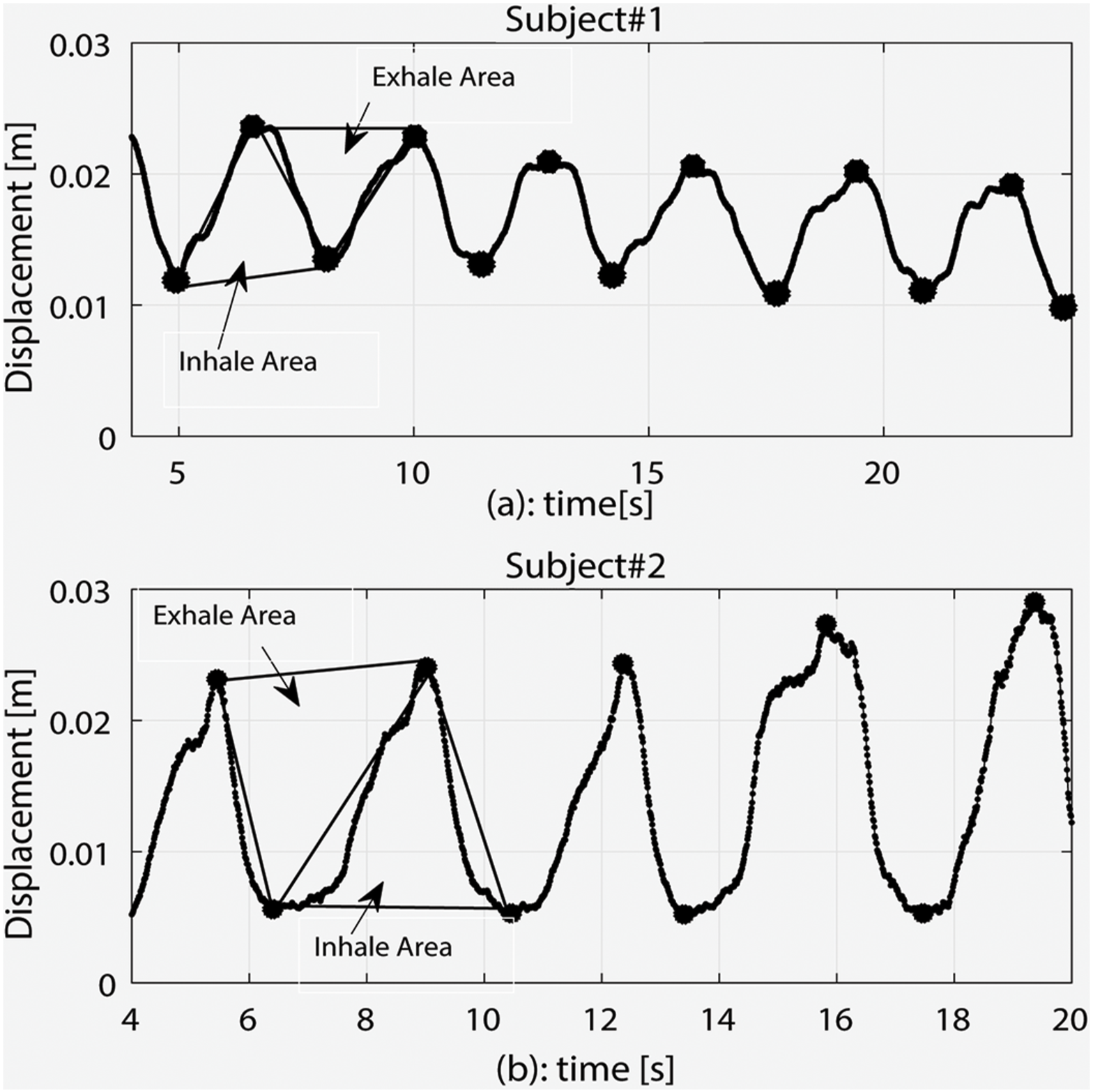}
\end{subfigure}
\hspace{\fill}
\begin{subfigure}[t]{0.24\textwidth}
\includegraphics[width=\textwidth,height=\textwidth]{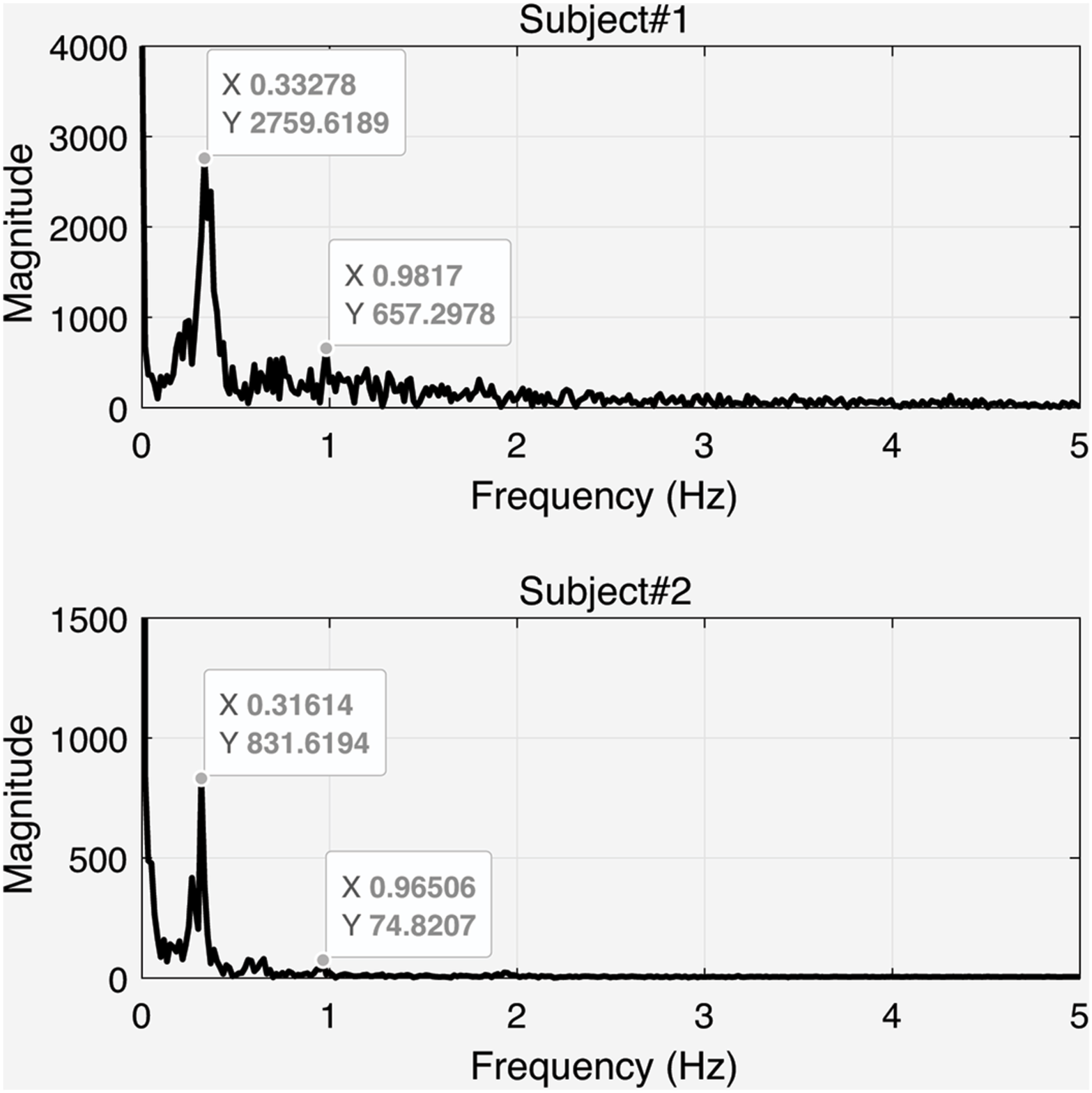}
\end{subfigure}

\caption{Left to right: (a) Raw IQ data received by different RF channels of the PRMS in Fig. \ref{fig:experiment}; (b) Breathing mixture obtained by (linear) demodulating the raw IQ (top), individual breathing patterns after source separation in comparison with the ones collected by the respiratory belt (bottom); (c,d) time and frequency domain analysis show the inhale and exhale characters are distinct between the two subjects, which allows patient tracking during modality changes.}
\label{fig:breathing separation}
\end{figure*}

\subsection{Implementing SIENNA}
SIENNA was developed and tested from 2019 to 2020, with two mmWave radars implemented for the preliminary system. The first design employs a commercial off-the-shelf 4-channel 24GHz monopulse radar transceiver (K-MC4 from RFbeam Microwave GmbH) interfacing a LabVIEW controlled DAQ through four LNA's (SR560 from Stanford Research System). The 3dB beam aperture is $12^{\circ}$ horizontally and $30^{\circ}$ vertically. The angle coverage is between +/-$15^{\circ}$ horizontally. The second design, aiming to address the reflection noise issue identified with the first design, is a customized 28GHz OFDM radar implemented via a mmWave beamforming development kit (BBox one/UD box from TMYTEK). The kit consists of a 16-channel 24-31GHz phased array antenna, and a frequency Up/Down converter, which interfaces with a LabVIEW controlled USRP (USRP-2974 from NI). The 3dB beam aperture is $13^{\circ}$ horizontally and $14^{\circ}$ vertically. The angle coverage is +/-$45^{\circ}$ horizontally and +/-$60^{\circ}$ vertically. 


The wireless respiratory belt is implemented with a piezo-electric respiration transducer (Model 1132 Pneumotrace II from UFI), which interfaces with a LabVIEW controlled DAQ. The Android application communicates via Bluetooth Low Energy (BLE) with the host computers that control the mmWave radar(s) and the respiration transducer to execute the modality-switching and data logging functionality. The implementation leverages the Android BLE APIs to connect to the Generic Attribute Profile (GATT) servers set up by the LabVIEW Bluetooth Toolkit on the host computers. The physical layer jammer for the chestband sensor is implemented via the Android application connecting to a Wi-Fi-BT-Bluetooth LE breakout board (ESP32-WROVER-IB from ESPRESSIF). The jammer for the mmWave radar is implemented via with a LabVIEW controlled USRP (USRP-2974 from NI).

\subsection{Experiment Setup}
We conducted laboratory and field experiments over a one-month period with 20 subjects selected through a random sample recruitment process. The subjects are between the ages of 16 and 35, weigh between 42 to 85 kilograms, and have no prior history of heart disease. The subjects are asked to participate in multiple trials of sleep studies under laboratory and everyday settings with environmental and adversarial complications detailed below.

\textbf{Sleep Environment and Data Collection.} We furnished two sleep environments for experimentation. The indoor laboratory environment consists of 2 twin size beds under a quiet and dim ambience. The outdoor field environment consists of two beach mats and umbrellas set up at a local beach park. During each trial, the subjects are attached with chestband sensors and positioned 0.5 meters below the mmWave radar system for data collection. We continuously monitor the respiration of the subjects for 1 hour, during which modality-switches and adversarial activities (if planned) are attempted every 10 minutes. After the experiment, we extract a 60-second data segment around each modality-switch to evaluate SIENNA's performance, with the remaining data to serve as references. Overall, we collected approximately 30,000 breathing cycle samples for each subject.



\textbf{Eavesdrop/Spoof.} We designed the eavesdropping and spoofing attacks with a BLE sniffer and spoofer, implemented via Ubertooth and Kismet \cite{technologies7010015}. During each experiment, one subject was asked to wear a chestband and lay down under a mmWave radar (Fig. \ref{fig:experiment}b). A third-party executed the modality switches and operated the computer running the Ubertooth. The packets transmitted by the OSA application, the chestband, and the mmWave radar were identified based on their Bluetooth Device Addresses (BDAs) obtained prior of the experiment. To implement the eavesdropping attack, the host's codes recorded the packets containing the fuzzy commitment and hash value of the new key during each modality-switch, which were analyzed offline in attempt to deduce the keys. To implement the spoofing attacks, an attacker-generated compliance tracking data encrypted with the deduced key was transmitted at higher power during data upload toward the OSA app, in attempt to manipulate the latter into accepting the fraudulent data, which was verified during offline analysis.

\subsection{Performance of Breathing Separation}
During each experiment, the signal (mixture) captured by the mmWave radar (Fig. \ref{fig:breathing separation}a) was filtered in real-time by the ICA-JADE-based breathing separation module implemented through a MATLAB script running within LabVIEW. The script filtered the signal with a digital FIR Low pass filter with cut off frequency of 10 Hz to suppress the high-frequency noise while preserving the physiological-related information. The filtered signal was linearly demodulated to compute the phase shifts caused by the displacement(s) of the chest(s) surface(s) during breathing (Fig. \ref{fig:breathing separation}b top). Specifically, the script calculated the covariance matrices of the I and Q channel signals and applied eigenvalue decomposition to the covariance matrices to extract the maximum chest displacement information. The demodulated signal was separated by the ICA-JADE method to isolate individual respiratory signatures (Fig. \ref{fig:breathing separation}b bottom). 

Finally, the script evaluated the performance of ICA-JADE by computing the cross-correlation between the isolated signatures with the ground truth obtained from the respiration transducer. The empirical result shows that the similarity is above 90\% when we limit the experiment within 60 seconds. As time increases, the subjects tend to move/turn on the bed, which changes the contribution of their breathing to the signal mixture, e.g., the mixing matrix, and the ICA-JADE result deviates from the truth. Thus, we limited the sensing period of SIENNA during the subsequent experiments to within 60 seconds to achieve a stable source separation.

\begin{figure*}[t]
\begin{subfigure}[t]{0.24\textwidth}
\includegraphics[width=\textwidth,height=\textwidth]{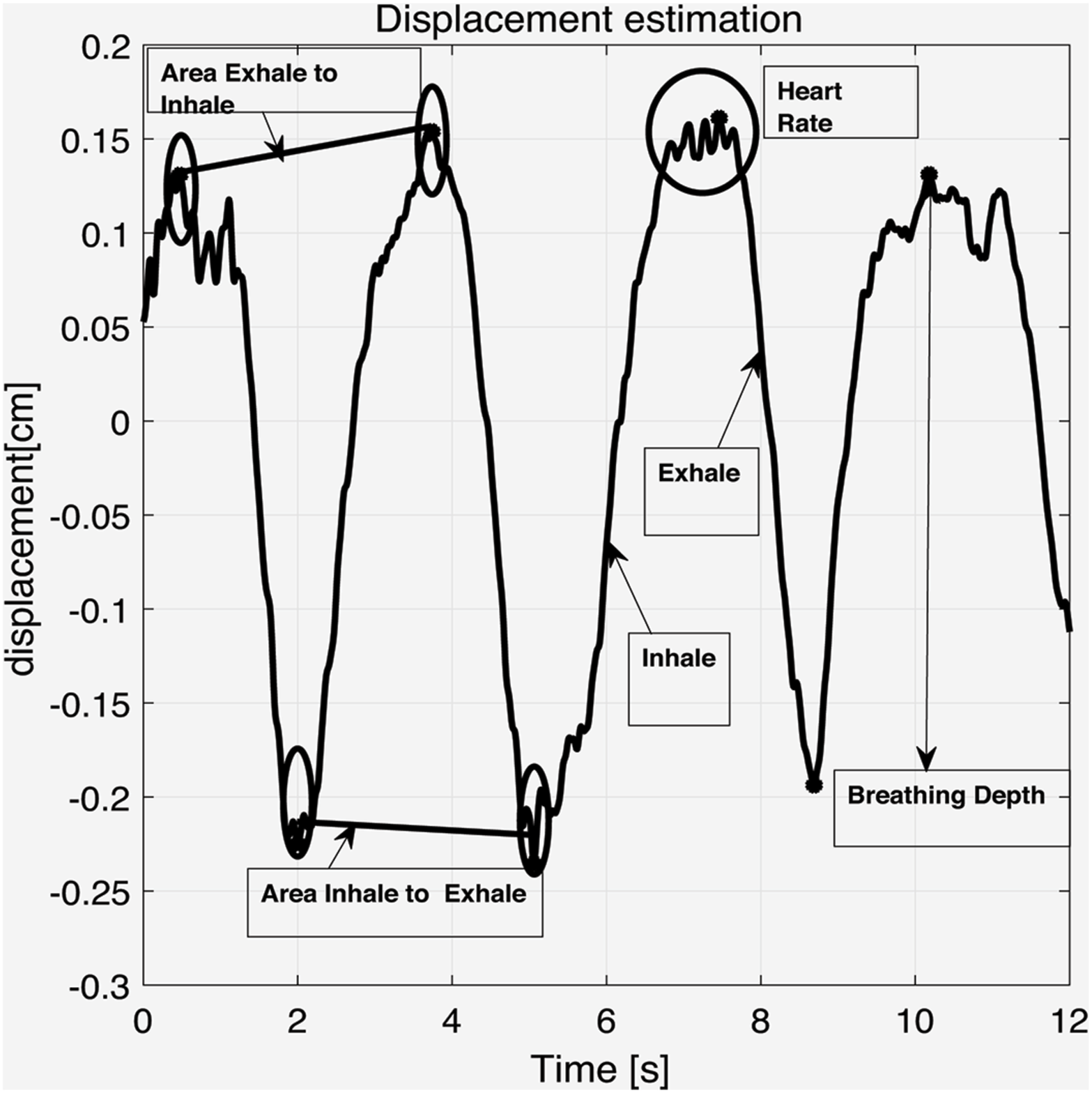}
\end{subfigure}
\hspace{\fill}
\begin{subfigure}[t]{0.24\textwidth}
\includegraphics[width=\textwidth]{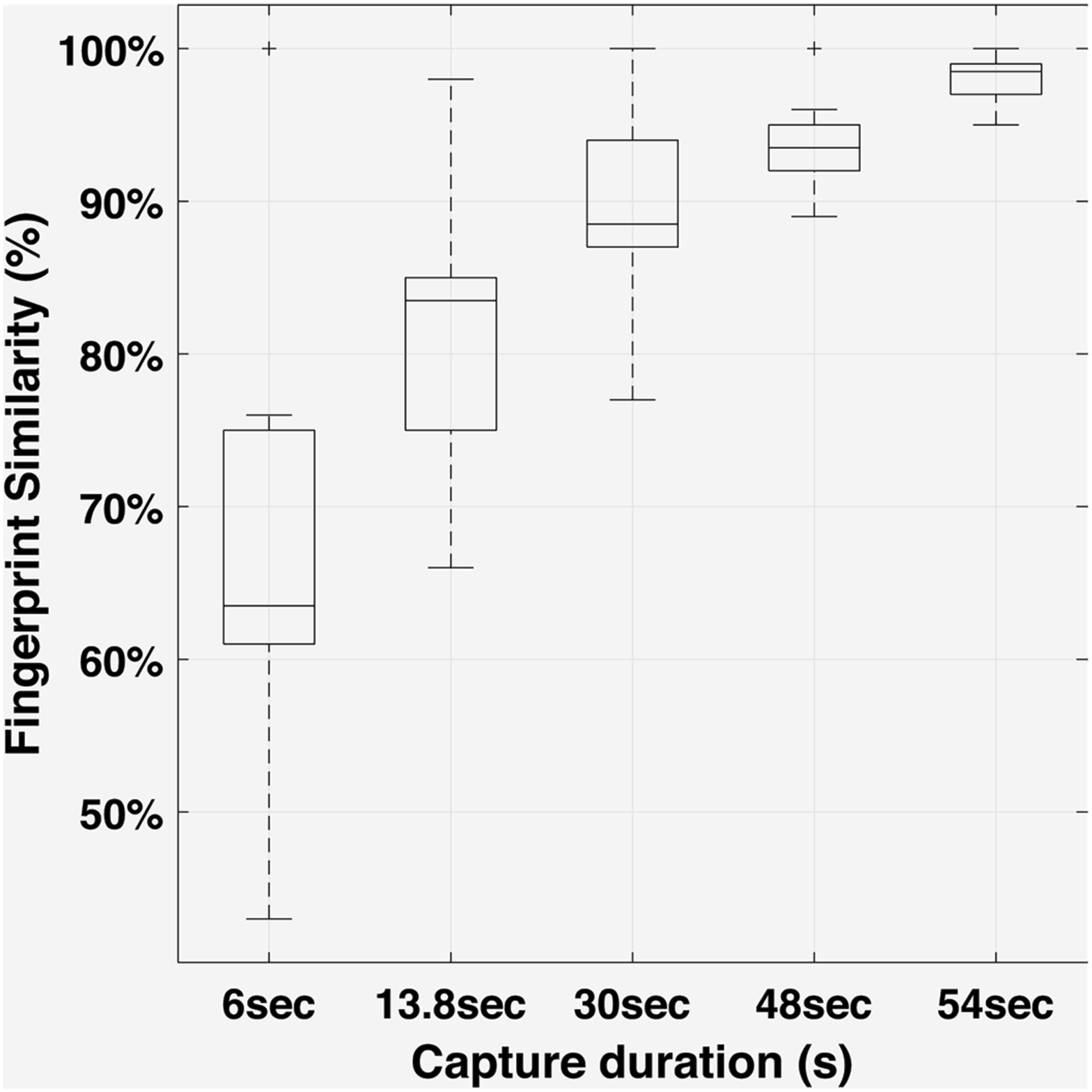}
\end{subfigure}
\hspace{\fill}
\begin{subfigure}[t]{0.24\textwidth}
\includegraphics[width=\textwidth,height=\textwidth]{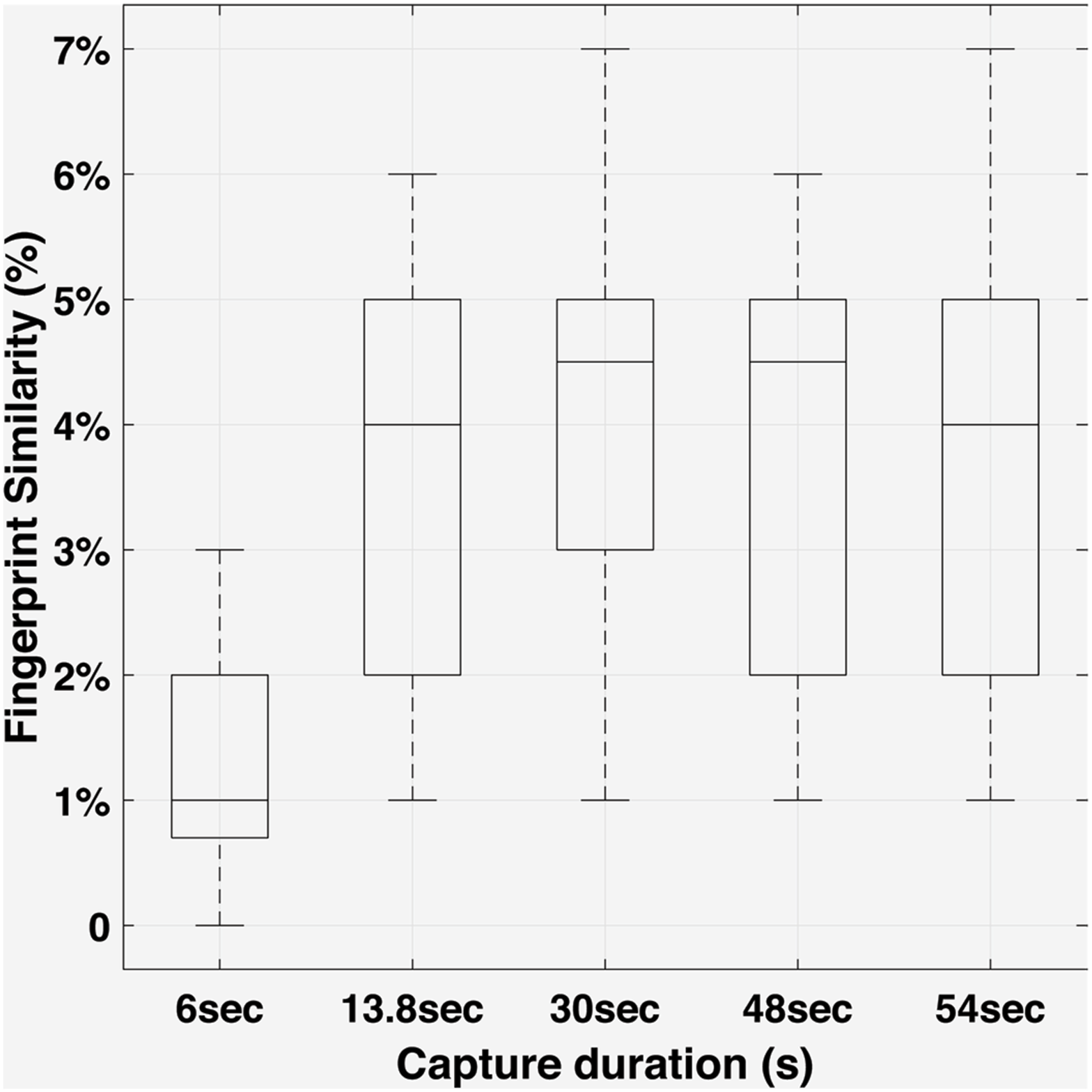}
\end{subfigure}
\hspace{\fill}
\begin{subfigure}[t]{0.24\textwidth}
\includegraphics[width=\textwidth,height=\textwidth]{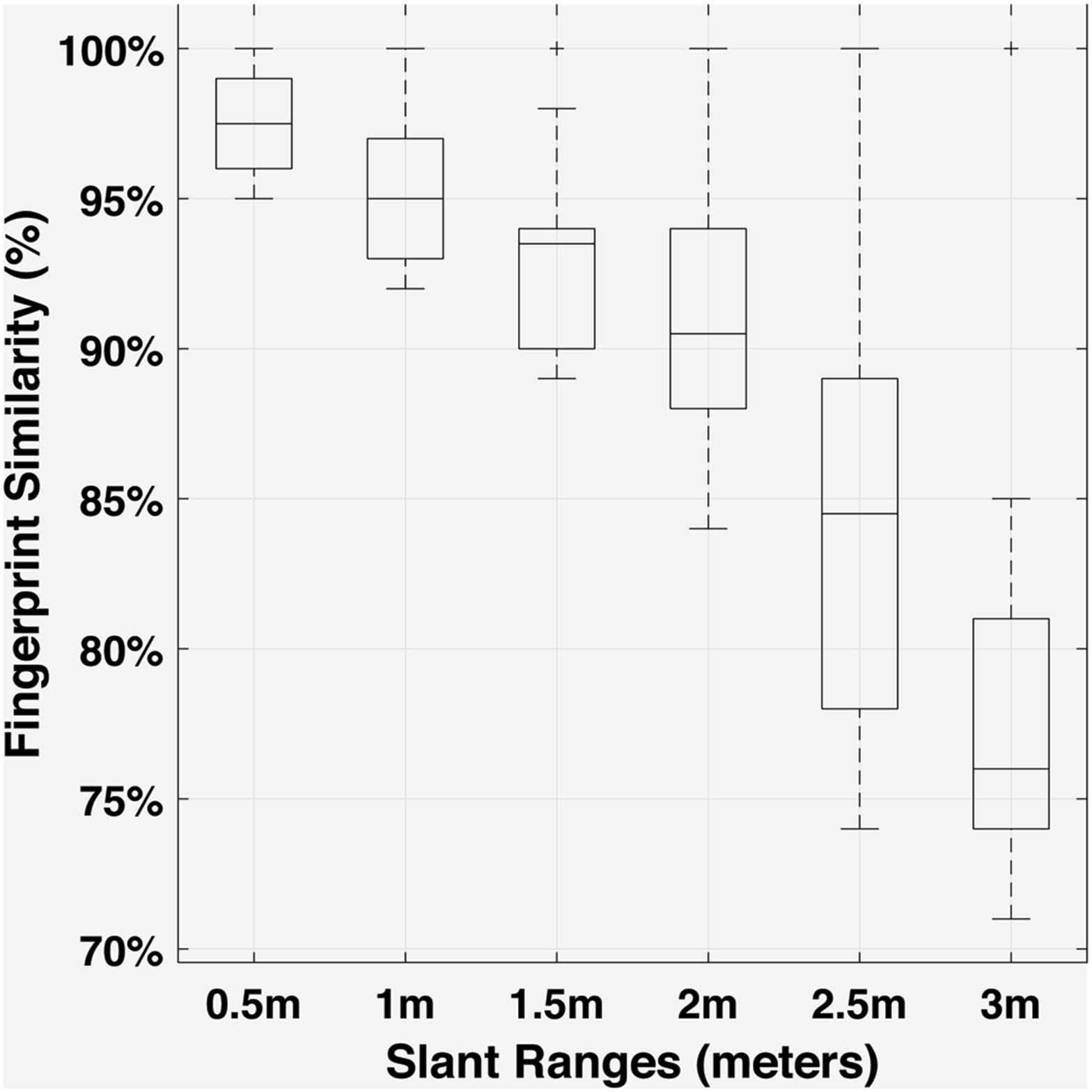}
\end{subfigure}
\caption{Left to right:(a) Signal reconstructed after 64-level-crossing quantization with vital related dynamic features preserved; (b)  Similarity between belt-based and PRMS-based breathing patterns, measured with the same subject; (c) Similarity between belt-based and PRMS-based breathing patterns, measured with different subjects; (d) The effect on the fingerprint similarity due to the change of slant range between the PRMS and the subject.}
\label{fig:fingerprint extraction}
\end{figure*}

\subsection{Performance of Fingerprint Extraction}
An individual breathing signature was quantized in parallel by multiple level-crossing LabVIEW VIs to generate the binary fingerprint after breathing separation. The breathing signature (torso deformation) due to changes in respiratory movement is a complex three-dimensional pattern, and varies greatly with subject parameters and activity context. Based on our preliminary data, the thorax motions due to respiration and heartbeat are limited within +/-0.5cm and +/-0.05cm (Fig. \ref{fig:fingerprint extraction}, and the rates of respiration and heartbeat are below 15 and 60 per minute, when the subject is at rest. Therefore, we employed ten level-crossing quantization branches with a quantization step size of 0.05cm at a sample rate of 10 per second, to preserve the fine-grained respiratory motion when extracting the breathing fingerprint.

To be compatible with the key evolution protocol, the quality of the binary fingerprints was evaluated based on the Hamming distances between fingerprints of the same subjects observed by different modalities, as well as different subjects observed by different modalities. It has been demonstrated in various works that human subjects can be sufficiently distinguished based on their inhales (local maxima), exhales (local minima), and breathing depth (the area between two consecutive maxima and in between one minima point). The similarities of these features directly translate to the Hamming distances between the quantized fingerprints. Therefore, comparing the Hamming distances is equivalent to constructing an equal-weighted linear classifier for patient identification.

The empirical results (Fig. \ref{fig:fingerprint extraction}b) show that the average Hamming similarity per bit between fingerprints of the same subject observed from different modalities is around 63\% when extracted from 6-second breathing signatures (with 2 breathing cycles). The Hamming similarity increases almost logarithmically towards 100\% as we extend the duration of the breathing signature to 60 seconds (with 20 breathing cycles). The elevated mean and reduced variance in the Hamming similarity is mostly due to the extended time, which allows repeated measurements of the periodic respiratory effort motion unique to each subject. 

On the contrary, the average Hamming similarity per bit between different subjects' fingerprints remains below 5\% despite the measurement duration (Fig. \ref{fig:fingerprint extraction}b). The position between the mmWave and the subject also poses a significant factor, which attenuates the Hamming similarity. As the subject moves away from the radar's downrange direction, the radar measurements deviate from the ground truth (obtained from the chestband). But their similarity remains sufficiently high compared to the breathing fingerprints between different subjects. Overall, the results show that the Hamming similarity per bit for SIENNA can be set to around 70\% to allow accurate patient tracking during modality switches.

\subsection{Performance of Key Evolution}
The binary breathing fingerprints, are chunked into multiple 10-second segments and padded with 0s to meet the codeword length of the $(2^{8}, 255, n)$ Reed-Solomon codes, e.g., $8 \times 255 = 2040$ bits. The particular group of Reed-Solomon codes are chosen to ensure the communicated data can be measured in exact multiple of bytes (8 bits). A key evolution salt of $8 \times n$ bits is randomly selected by the on-duty sensor and encoded with the $\textsc{RS}(2^{8}, 255, n)$ Reed-Solomon encoder to generate the 2040 bits opening value. The opening value and the (multiple) padded fingerprint segments are XORed together to generate the 2040 bits commitment.

\begin{figure*}[t]
\begin{subfigure}[t]{0.24\textwidth}
\includegraphics[width=\textwidth,height=\textwidth]{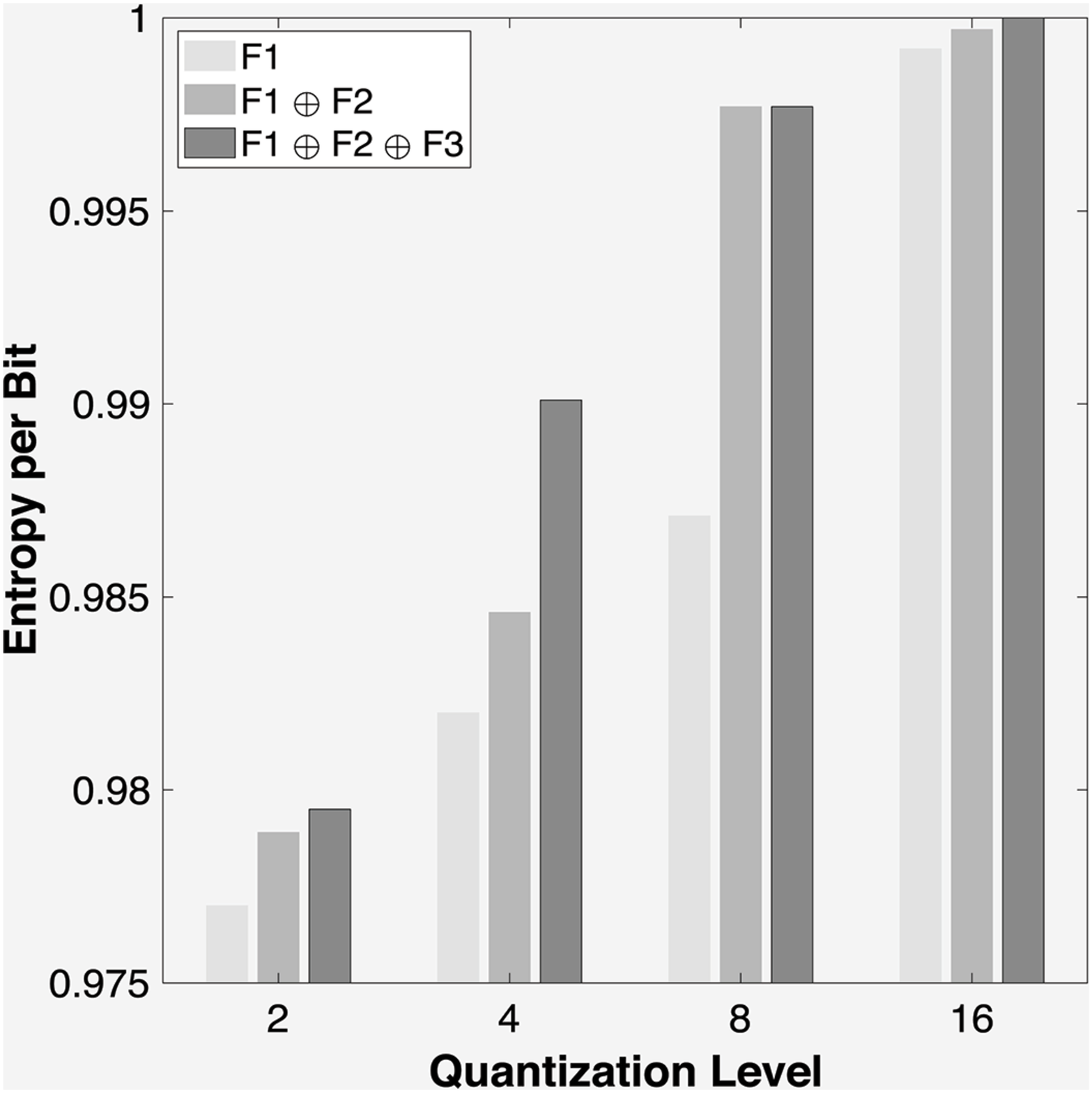}
\end{subfigure}
\hspace{\fill}
\begin{subfigure}[t]{0.24\textwidth}
\includegraphics[width=\textwidth,height=\textwidth]{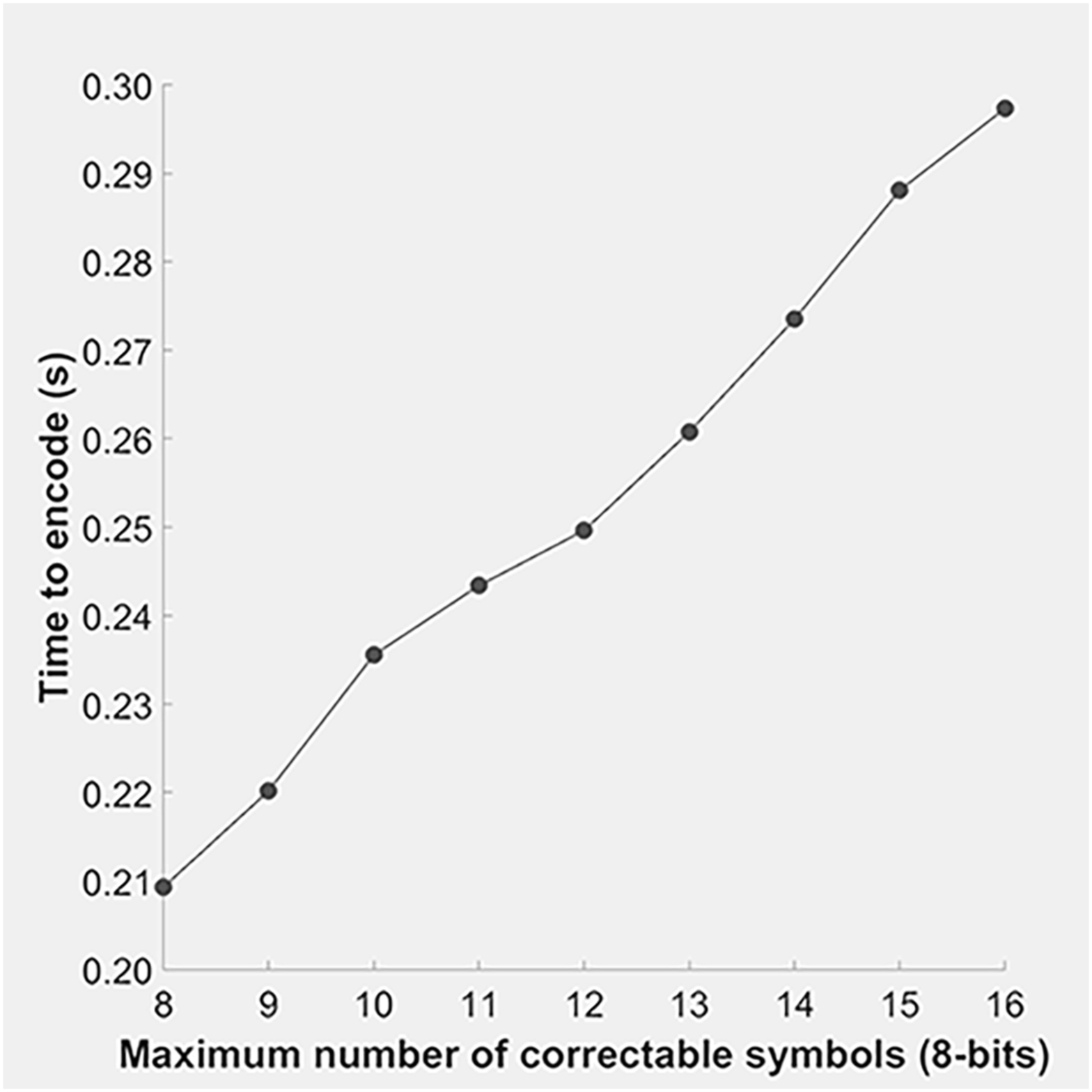}
\end{subfigure}
\hspace{\fill}
\begin{subfigure}[t]{0.24\textwidth}
\includegraphics[width=\textwidth,height=\textwidth]{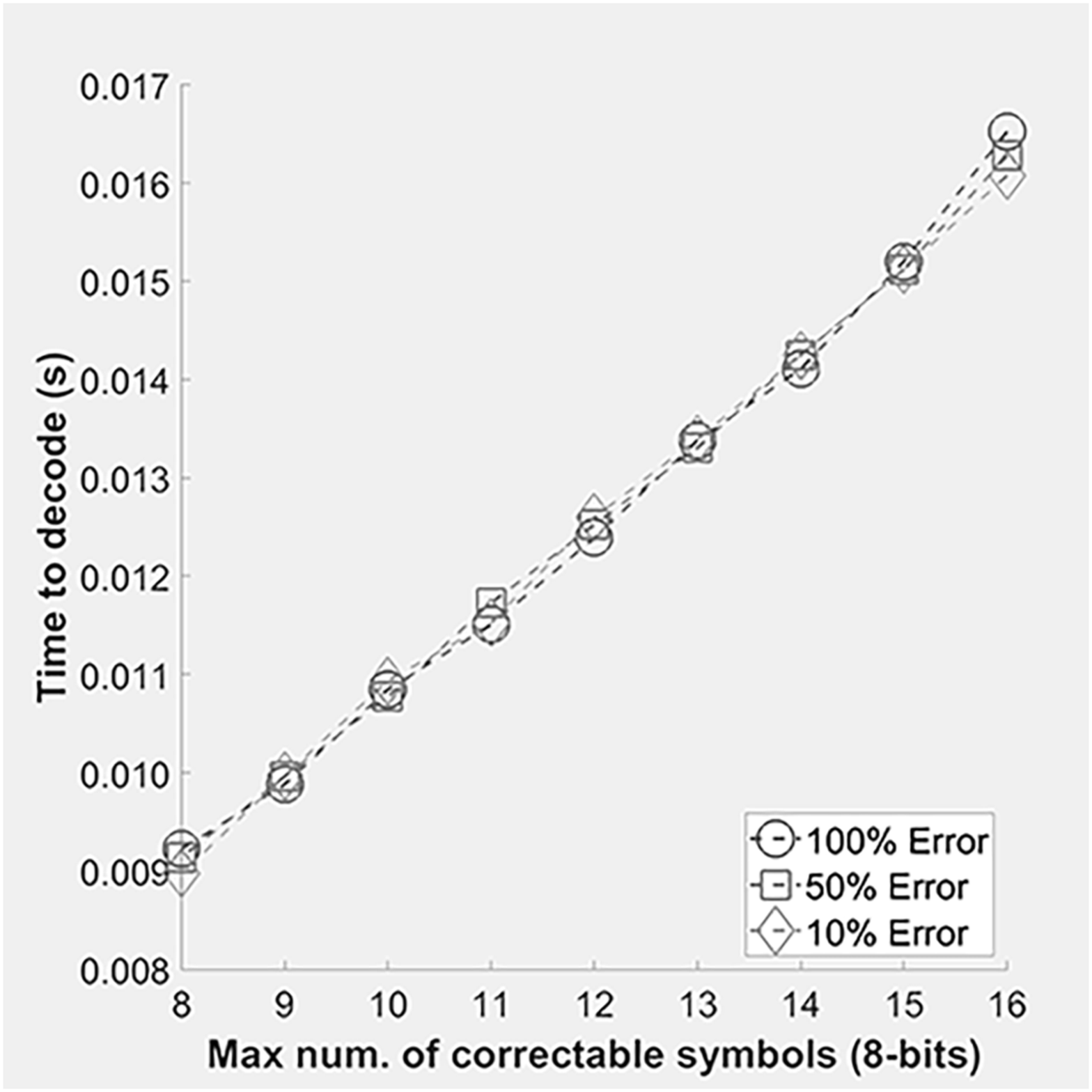}
\end{subfigure}
\hspace{\fill}
\begin{subfigure}[t]{0.24\textwidth}
\includegraphics[width=\textwidth,height=\textwidth]{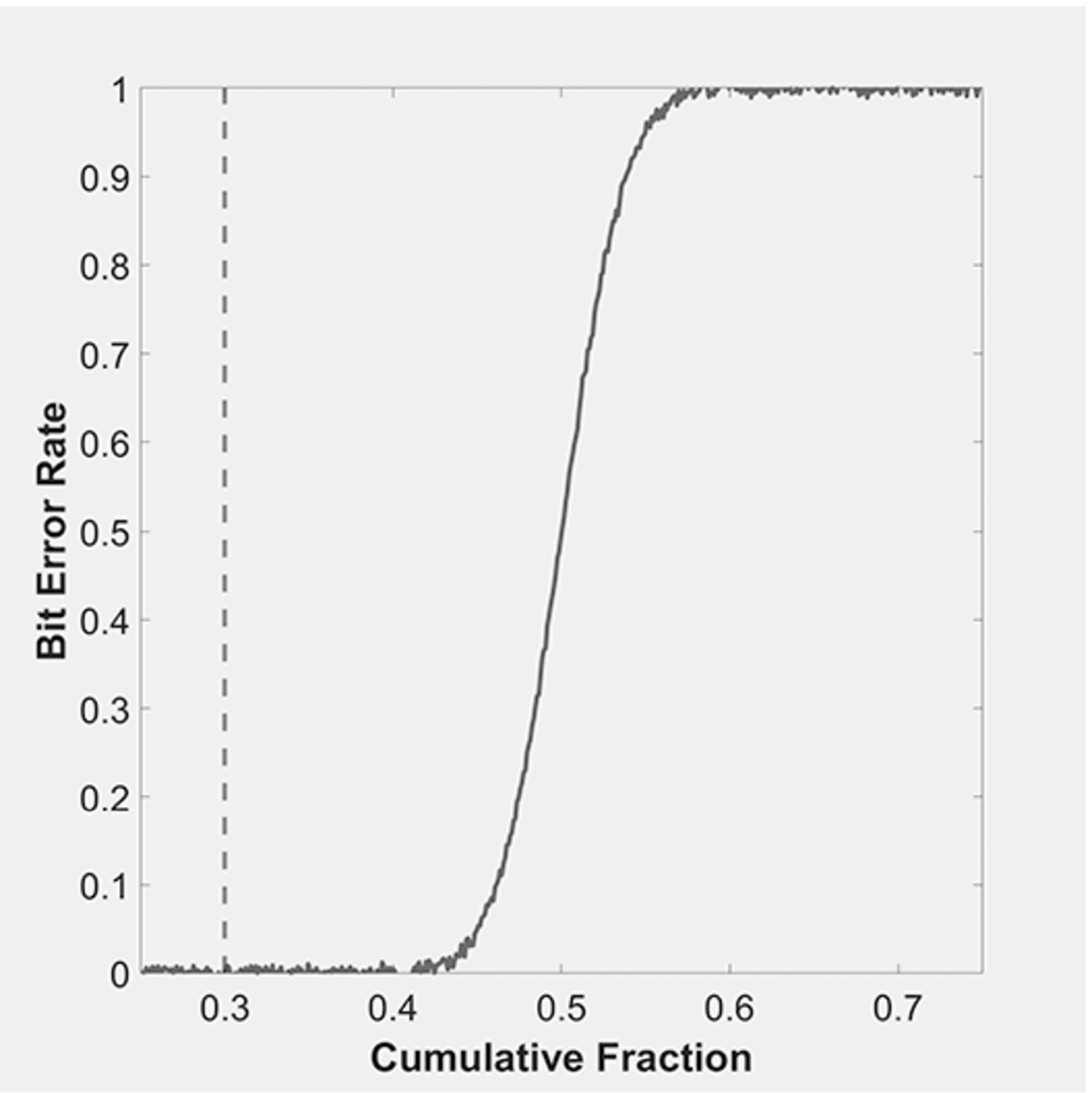}
\end{subfigure}
\caption{Left to right: (a) Average entropy per bit of the fuzzy commitments, e.g., RS encoded key salt XORed with (multiple) fingerprint segments, measured via NIST tests; (b) Commitment time with RS codes of different message/parity symbol lengths; (c) Reconstruction time with RS codes of different message/parity symbol lengths and symbols errors; (d) Performance of SIENNA against eavesdropping and spoofing in terms of aggregated
BER.}
\label{fig:key evolution}
\end{figure*}

We evaluate the security of the fuzzy commitment via a randomness test using the NIST statistical test suite. Based on 10 million randomly generated key evolution salts (with entropy per bit equal to one), we measured the randomness of the opening values and the commitments (Fig. \ref{fig:key evolution}a). The empirical test shows that the entropy per bit drops almost by half when the key salt is converted into a commitment. In other words, the entropy of a 2040-bit commitment is approximately 1000 bits. The primary causes of the reduction are due to the redundancy in the Reed-Solomon codes and the human respiratory motion's cyclic character. Two additional factors that affect the randomness of the commitment are the quantization levels and the number of the XOR operation rounds. When the quantization levels increase, the granularity of the binary sequencing improves, slightly elevating the randomness of the breathing fingerprints, resulting in a higher degree of entropy in the commitments. When the commitment is generated with multiple rounds of XOR operations, the entropy decreases due to the cross-correlation between fingerprint segments\footnote{In an XOR operation, the higher correlation between the inputs, the less the entropy in the output.}.

We further measured the commitment and reconstruction time of the key evolution protocol. As the binary quantization takes negligible time to perform, the Reed-Solomon encoder and decoder's efficiency, besides communication delay,  dominates the protocol's turnaround time. We timed our LabVIEW-based protocol implementation with different parity symbol lengths, $k = 255 - n$, in the $(2^{8}, 255, n)$ Reed-Solomon codes. The results show that the overall commitment time is below 0.3 seconds and grows linearly to the maximum number of correctable symbols, $k/2$, due to the additional Lagrange interpolations needed to compute the parity symbols (Fig. \ref{fig:key evolution}b). The overall reconstruction time is below 0.02 seconds and grows linearly to $k/2$, due to the additional syndromes computed for error corrections (Fig. \ref{fig:key evolution}c). The decoding time is invariance to the number of errors in the codewords, which is a security advantage as the attacker cannot infer the decoded message's correctness based on decoding time.

\vspace{-10pt}

\subsection{Performance under Adversarial Settings}


SIENNA's performance against eavesdropping and spoofing is evaluated by comparing the aggregated bit error rate (BER) at the receiver versus the aggregated BER at the attacker side. Due to the application of fuzzy commitment, the key establishment protocol allows a maximum of 27\% BER in the breathing fingerprints (when using $(2^8, 255, 201)$ Reed-Solomon codes) to recover the key salt. Compared to Fig. \ref{fig:fingerprint extraction}c, such a BER alone prevents any outside attackers who cannot observe and mimick the patient's breathing patterns from stealing the key salt. Our experiment further showed that the jamming signal could suppress the attacker's BER to approximately 50\% within the PRMS's transmission range, rendering the fuzzy commitment message that is not decodable, even if an attacker could obtain the patient's breathing fingerprint. Overall, the CDF of the accumulated BERs for attackers at any locations within the PRMS's transmission range concentrated between 41\% to 50\% (Fig. \ref{fig:key evolution}d) and is well beyond the correctable range of the selected Reed-Solomon codes. Thus, the combination of both techniques ensures sufficient and ensures complete protection of the security key during modality-switching against both outsider and insider attacks.




\section{Conclusion}
\label{sec:conclusion}
We presented SIENNA, a novel insider-resistant context-based pairing scheme for multi-modality OSA screening systems. By merging fuzzy commitment, friendly jamming, and JADE-ICA, SIENNA leverages the unique patterns of a person's breathing dynamics for secure pairing with the presence of co-located attackers. We formally analyzed the security of SIENNA according to the attacker's knowledge of the extracted binary sequence. Our results show that the combination of fuzzy commitment, friendly jamming, and JADE-ICA in SIENNA can protect the security key during the pairing process against an attacker equipped with complete knowledge of the context information, and is robust within a noisy at-home environment with multiple persons. In the future, we plan to further optimize SIENNA in terms of power consumption and execution time.


\ifCLASSOPTIONcaptionsoff
\newpage
\fi

\printbibliography[title=References]

\end{document}